\documentclass[aps,prd,nofootinbib,onecolumn,notitlepage]{revtex4-1}
\usepackage[utf8]{inputenc}
\usepackage{graphicx}
\usepackage{amsmath}
\usepackage{amsfonts}
\usepackage{amssymb}
\usepackage{natbib}
\usepackage{color}
\usepackage[unicode=true,pdfusetitle,
 bookmarks=true,bookmarksnumbered=false,bookmarksopen=false,
 breaklinks=false,backref=false,colorlinks=false]
 {hyperref}
\usepackage{bbm}
\usepackage{amsfonts}
\usepackage{amssymb}
\usepackage{xcolor}
\usepackage[normalem]{ulem}

\begin{document}
\title{Hawking radiation and black hole gravitational
back reaction - A quantum geometrodynamical simplified model}
\author{Jo\~ao Marto$^{1,2}$}
\affiliation{$^{1}$Departamento de F\'isica, Universidade da Beira Interior, Rua Marquês D'Ávila e Bolama, 6201-001 Covilhã, Portugal.}
\affiliation{$^{2}$Centro de Matemática e Aplicações da Universidade da Beira Interior
(CMA-UBI), Rua Marquês D'Ávila e Bolama, 6201-001 Covilhã, Portugal.}
\date{\today}
\begin{abstract}
The purpose of this paper is to analyse the back reaction problem,
between Hawking radiation and the black hole, in a simplified model
for the black hole evaporation in the quantum geometrodynamics context. 
The idea is to transcribe the most important
characteristics of the Wheeler-DeWitt equation into a Schrödinger's
type of equation. Subsequently, we consider Hawking radiation and
black hole quantum states evolution under the influence of a potential
that includes back reaction. Finally, entropy is estimated as a measure
of the entanglement between the black hole and Hawking radiation states
in this model.
\end{abstract}
 \vskip -1.0 truecm

\maketitle

\section{Introduction}
Since the discovery that black holes would have to emit radiation,
there have been proposals to explain the loss of information associated
with the apparent conversion of pure to mixed quantum states. From
the beginning, this information loss was proposed to be fundamental
and, the non unitary evolution of pure to mixed quantum states constituted
a hypothesis to solve the problem associated with this loss. For example,
Steven Hawking own proposal of the non unitary evolution is represented
by the ``dollar matrix'' $S\hspace{-7pt}/$ \cite{Hawking1976}
\[
\rho^{\mathrm{final}}=S\hspace{-7pt}/\rho^{\mathrm{initial}}\:,
\]
which allows the evolution of pure quantum states, characterised by
the density matrix $\rho^{\mathrm{initial}}$, into mixed states $\rho^{\mathrm{final}}$.

The black hole evaporation mechanism and the problem of information
loss, collected behind the event horizon, constituted a privileged
arena for quantum gravity theories candidates (namely, quantum geometrodynamics
\cite{Kiefer2007}, string theory \cite{Ramond_v1,Neveu_v1,Scherk_v1,becker_becker_schwarz_2006,Zwiebach:2004tj} and loop quantum gravity \cite{Ashtekar_v1,rovelli_2004,thiemann_2007,bojowald_2010}) to establish themselves beyond General Relativity.
However, the scientific community was reluctant to give up unitarity,
a crucial feature of Quantum Mechanics, and the hypothesis of a new
principle of complementarity, between the points of views of an infinitely
distant observer and a free falling observer near the event horizon,
was raised \cite{Susskind:1993if}. Following a similar approach,
it has been emphasised over the time the role of the gravitational
back reaction effect \cite{THOOFT1985727,DRAY1985173} of Hawking
radiation on the event horizon as a way to allow the information accumulated
within the black hole to be encoded in the outgoing radiation. In
this way, the emergence of a mechanism in which all the black hole
information (a four dimensional object in General Relativity) would
be accessible at the event horizon (which can be described as a membrane
with one dimension less than the black hole), is somehow similar to
what happens with a hologram \cite{tHooft:1993dmi}. This new holographic
principle was simultaneously proposed and clarified
\cite{Susskind:1994vu,Susskind_book2004} in order to incorporate
the aforementioned principle of complementarity. The next step happened
when it was conjectured the correspondence between
classes of quantum gravity theories (5-dimensional anti-De Sitter
solutions in string theories) and conformal field theory (CFT - conformal
field theory - 4-dimensional boundary of the 5-dimensional solutions),
the so called AdS/CFT conjecture \cite{Maldacena:1997re,Aharony:1999ti,Natsuume:2014sfa,nastase_2015}.
This discovery was extremely important to ensure the possibility of
a correspondence between the physics that describes the interior of
the black hole (supposedly quantum gravity), and the existence of
a quantum field theory at its boundary (the surface that defines the
event horizon) that would allow to save the unitarity.

In 2012, in an effort to analyse important assumptions, such as: 1)
the principle of complementarity proposed by Susskind and its colaborators,
2) the AdS/CFT correspondence and, 3) the equivalence principle of
General Relativity, in the way that Hawking radiation could encode
the information stored in the black hole, a new paradox was discovered
\cite{Almheiri2013}. In simple terms, the impossibility of having
the particle, which leaves the black hole, in an maximally entangled state (or
non factored state) with two systems simultaneously (the pair disappearing
beyond the horizon and all the Hawking radiation emitted in the past, a problem related to the so-called monogamy of entanglement),
leads to postulate the existence of a firewall that would destroy
any free falling observer trying to cross the event horizon. The firewall
existence is incompatible with Einstein's equivalence principle. However,
if there is no firewall, and the principle of equivalence is respected,
according to these authors, unitarity is
lost and information loss is inevitable. Apparently, the situation
is such that either General Relativity principles or Quantum Mechanics
principles need to be reviewed \cite{Polchinski:2016hrw,Harlow:2014yka}.
This is an open problem and the role of gravitational back reaction,
between Hawking radiation and the black hole, persists as an unknown
and potentially enlightening mechanism on how to correctly formulate
a quantum theory of the gravitational field.

In an attempt to study the possible gravitational back reaction,
between Hawking radiation and the black hole, from the quantum geometrodynamics point of view, a toy model was proposed \citep{Kiefer:2008av}. It was shown and discussed the conditions under which the Wheeler-DeWitt equation could be used to describe a quantum black hole. In particular, a simple model for the black hole evaporation was studied using a Schrödinger type of equation and, the cases for initial squeezed ground states and coherent states were taken to represent the initial black hole quantum state. One can ask, how can a complex equation such as the Wheeler-DeWitt be approximated by a Schrödinger type of equation?
In the cosmological context, several formal derivations were carried   \cite{Kiefer2007,Lapchinsky1979fd,HalliwellandHawking1985,BANKS1985,BARVINSKY1989}
 with the purpose of enabling to use the limit of a quantum field theory in an external space-time for the full quantum gravity theory. Such approaches usually involve  procedures like the Born-Oppenheimer or Wentzel-Kramers-Brillouin (WKB) approximations.
 
In this work we review this toy model. It is important to notice that, even though, a full study of the time evolution of the Hawking radiation and black hole quantum states was performed when a simple back reaction term is introduced, an important part of the discussion about the time evolution of the resulting entangled state was left incomplete. In fact, it is exactly the motivation of this paper to address the problem of explicitly describe the time evolution of the degree of entanglement of this quantum system. In addition, another important goal is to get an approximate estimate of the Von Neumann entropy and check is the back reaction can induce a release of the quantum information in the Hawking radiation. These results can be interesting, in the quantum geometrodynamics context, as a simple starting point to more robustly address the black hole information paradox in a canonical quantization of gravity program.

This paper is organized as follow. In sections \ref{section2} and \ref{section3} we present an introduction to quantum geometrodynamics and consider a semiclassical approximation of the Wheeler-DeWitt equation. In sections \ref{section4} and \ref{section5} we derive a simple model of the back reaction between the Hawking radiation and the  black hole quantum states, where its dynamics is governed by a Schrödinger type of equation. Finally, in sections \ref{section6} and \ref{section7}, we obtain and discuss the main result of this paper, namely the time evolution of the entanglement entropy and the behaviour of the quantum information of the Hawking radiation state.
 
\section{Quantum geometrodynamics and the semiclassical approximation}
\label{section2}
In the following, we mention a brief description of the bases of J.A.
Wheeler's geometrodynamics, which consists in a 3+1 spacetime
decomposition (ADM decomposition - R. Arnowitt, S. Deser and C.W.
Misner \cite{Arnowitt}), and obtain General Relativity field equations in that
context. The field equations, obtained in this procedure, will exhibit
the evolution of a pair of dynamical variables $\left(h_{ab},\,K_{ab}\right)$
- the 3-dimensional metric $h_{ab}$ (induced metric) and the extrinsic
curvature $K_{ab}$ - on a Cauchy hypersurface $\Sigma_{t}$ (three-dimensional
surface).

General Relativity, defined by the Einstein-Hilbert action\footnote{Without the cosmological constant.}
\begin{equation}
S_{\mathrm{EH}}=\dfrac{c^{4}}{16\pi G}\int d^{4}x\:\sqrt{-g}\,R\:,\label{Einstein Hilbert action}
\end{equation}
can be expressed under the hamiltonian formalism. For this purpose,
a $3+1$ decomposition of spacetime $\left(\mathcal{M},\,g\right)$\footnote{We assume that this spacetime is globally hyperbolic, such that we
ensure that it can be foliated in Cauchy hypersurfaces.} may be considered, where $\mathcal{M}$ is a smooth manifold and
$g$ lorentzian metric in $\mathcal{M}$ . Moreover, this decomposition
consists of the 4-dimensional spacetime foliation in a continuous
sequence of Cauchy hypersurfaces $\Sigma_{t}$, parameterised by a
global time variable $t$\footnote{For which a flow of `time' can be perceived when a observer world
line crosses a sequence of Cauchy hypersurfaces. }. General Relativity covariance is maintained, in this procedure,
by considering all possible ways of carrying this foliation. When
we consider the hamiltonian formalism we need to define a pair of
canonical variables, however, we can initially identify a pair of
dynamical variables constituted, on the one hand, by the 3-dimensional
metric induced in $\Sigma_{t}$ by the spacetime metric
\begin{equation}
\mathbf{h=g+n\otimes n}\quad\left(h_{\mu\nu}=g_{\mu\nu}+n_{\mu}n_{\nu}\right)\:,\label{3D-spatial metric}
\end{equation}
where $n_{\mu}$ is an ortogonal vector to $\Sigma_{t}$. In this
way we can separate the metric $g$, in its temporal e spatial components,
according to the following expressions,
\begin{equation}
g_{\mu\nu}=\left(\begin{array}{cc}
N_{a}N^{a}-N^{2} & N_{b}\\
N_{c} & h_{ab}
\end{array}\right)\quad\mathrm{and}\quad g^{\mu\nu}=\left(\begin{array}{cc}
-\dfrac{1}{N^{2}} & \dfrac{N^{b}}{N^{2}}\\
\dfrac{N^{c}}{N2} & h_{ab}-\dfrac{N^{a}N^{b}}{N^{2}}
\end{array}\right),\label{g-metric foliation}
\end{equation}
or, in a more suitable compact form,
\begin{equation}
g_{\mu\nu}dx^{\mu}dx^{\nu}=-N^{2}dt^{2}+h_{ab}\left(dx^{a}+N^{a}dt\right)\left(dx^{b}+N^{b}dt\right)\:.
\end{equation}
In the previous equation $N$ is called the lapse function whereas
$N^{a}$ is the shift vector. The other canonical variable, on the
other hand, is the extrinsic curvature
\begin{equation}
K_{\mu\nu}=h_{\mu}^{\sigma}\nabla_{\sigma}n_{\nu}\:.\label{Extrinsic curvature}
\end{equation}
Hence, the dynamical variables pair $\left(h_{ab},\,K_{ab}\right)$
(with Latin letter indexes, defining 3-dimensional tensor fields)
enable us to rewrite Einstein-Hilbert action (\ref{Einstein Hilbert action})
as
\begin{equation}
S_{\mathrm{EH}}=\dfrac{c^{4}}{16\pi G}\int_{\mathcal{M}}dtd^{3}x\:\mathcal{L}=\dfrac{c^{4}}{16\pi G}\int_{\mathcal{M}}dtd^{3}x\:N\sqrt{h}\left(K_{ab}K^{ab}-K^{2}+^{\left(3\right)}R\right)\:.\label{Hamiltonian action}
\end{equation}
We can notice that the lapse function and the shift vector are Lagrange
multipliers (since $\partial\mathcal{L}/\partial\dot{N}=0$ e $\partial\mathcal{L}/\partial\dot{N}_{a}=0$
) and, according to Dirac \cite{Dirac:1964:LQM} we can establish
the existence of primary constraints which allow to write the action
(\ref{Hamiltonian action}) as
\begin{equation}
S_{\mathrm{EH}}=\dfrac{c^{4}}{16\pi G}\int_{\mathcal{\mathcal{M}}}dtd^{3}x\:\left(p^{ab}\dot{h}_{ab}-N\mathcal{H}_{\perp}^{g}-N^{a}\mathcal{H}_{a}^{g}\right)\:,\label{Action with multipliers}
\end{equation}
with $p^{ab}=\partial\mathcal{L}/\partial\dot{h}_{ab}$ (conjugate
momentum of the dynamical variable $h_{ab}$) and where
\begin{equation}
\begin{cases}
\mathcal{H}_{\perp}^{g} & =\dfrac{16\pi G}{c^{4}}\,G_{abcd}p^{ab}p^{cd}-\dfrac{c^{4}\sqrt{h}}{16\pi G}{}^{\left(3\right)}R\\
\mathcal{H}_{a}^{g} & =-2D_{b}\left(h_{ac}p^{bc}\right)
\end{cases}\:,\label{diffeo-constr}
\end{equation}
with $G_{abcd}=\dfrac{1}{2\sqrt{h}}\left(h_{ac}h_{bd}+h_{ad}h_{bc}-h_{ab}h_{cd}\right)$
being the DeWitt metric and $D_{b}$ is the covariant derivative.
We can define the hamiltonian constraint
\begin{equation}
\mathcal{H}_{\perp}^{g}\approx0\:,\label{Hamiltonian constraint}
\end{equation}
and the diffeomorphism constraint
\begin{equation}
\mathcal{H}_{a}^{g}\approx0\:,\label{Diffeomorphism constraint}
\end{equation}
through the variation of the action (\ref{Action with multipliers})
with respect to $N$ and $N^{a}$. Physically, constraints (\ref{Hamiltonian constraint})-(\ref{Diffeomorphism constraint})
express the freedom to choose any coordinate system in General Relativity.
More precisely, the choice of the particular foliation $\Sigma_{t}$
is equivalent to choose the lapse function $N$ and, the spatial coordinates
$\left(x^{i}\right)$ choice is equivalent to choose a particular
shift vector $N^{a}$. It is important to emphasise that, related
to the DeWitt metric $G_{abcd}$ definition, the kinetic term in equation
(\ref{Hamiltonian constraint}) is indefinite, since not all kinetic functional operators in Eq. \ref{diffeo-constr} share the same sign. 
This property will
persist beyond the quantisation procedure and will play a crucial
role in the semiclassical (where it will give rise to a negative kinetic
term) approach to the black hole evaporation process.

\section{Canonical variables quantisation and Wheeler-DeWitt equation}
\label{section3}
The canonical quantisation programme, according to P.M. Dirac prescription,
demands the transition of classical to quantum canonical variables
$\left(h_{ab},\,p_{ab}\right)\rightarrow\left(\hat{h}_{ab},\,-i\hbar\delta/\delta h^{ab}\right)$,
and also promotes Poisson brackets to commutators. We have to define
a wave state functional $\Psi\left(h_{ab}\right)$ belonging to the
space of all 3-dimensional metrics Riem $\Sigma$. Nevertheless, there
are important issues related with:
\begin{enumerate}
\item the correct factor ordering in building quantum observables from the
fundamental variables $\left(\hat{h}_{ab},\,-i\hbar\delta/\delta h^{ab}\right)$,
\item the interpretation of quantum observables as operators acting on the
wave functional $\Psi\left(h_{ab}\right)$ and the adequate definition
of a Hilbert space,
\item the classical constraints (\ref{Hamiltonian constraint})-(\ref{Diffeomorphism constraint})
conversion to their quantum counterpart 
\begin{equation}
\begin{cases}
\mathcal{H}_{\perp}^{g}\Psi\left(h_{ab}\right) & =0\\
\mathcal{H}_{a}^{g}\Psi\left(h_{ab}\right) & =0
\end{cases}\:,\label{Hamil-Quantum constraints}
\end{equation}
and their quantum interpretation,
\item the lack of time evolution in the previous quantum constraints.
\end{enumerate}
These questions are thoroughly discussed in \cite{Kiefer2007,Kiefer:2008sw},
as well as possible solutions and open problems till the present day.
Among the previous mentioned issues, the problem related to the lack
of time evolution seems to stand as an essential feature in the formulation
of a quantum theory of the gravitational field. If we assume that
the wave functional evolution over time depends on a time concept
defined after the canonical quantisation, then, the time parameter
$t$ will be an emergent quantity \cite{Kiefer:2013jqa}. 

In order to address the black hole evaporation problem and, to explore
how information is eventually encoded in Hawking radiation, it would
be important to obtain the entropy time evolution as a measure of
the degree of quantum entanglement between radiation and black hole
states. Since the quantum version of the hamiltonian constraint (\ref{Hamiltonian constraint}),
\begin{equation}
\mathcal{H}_{\perp}^{g}\Psi\left(h_{ab}\right)=\left(\dfrac{16\pi G\hbar^{2}}{c^{4}}\,G_{abcd}\dfrac{\delta^{2}}{\delta h_{ab}\delta h_{cd}}-\dfrac{c^{4}\sqrt{h}}{16\pi G}{}^{\left(3\right)}R\right)\Psi\left(h_{ab}\right)=0\:,\label{Wheeler-De Witt Eq}
\end{equation}
known as Wheeler-DeWitt equation, and the quantum diffeomorphism constraint
\begin{equation}
\mathcal{H}_{a}^{g}\Psi\left(h_{ab}\right)=D_{b}\left(h_{ac}\dfrac{\delta}{\delta h_{bc}}\right)\Psi\left(h_{ab}\right)=0\label{Diffeo quantum Eq}
\end{equation}
are both time independent, the wave functional is connected to a purely
quantum and closed gravitational system. In the case involving the
study of a black hole evaporation phase, equations (\ref{Wheeler-De Witt Eq})-(\ref{Diffeo quantum Eq})
describe a quantum black hole in the context of a purely quantum universe.
This situation is not suitable if we consider that we must have several
classical observers measuring and depicting the time evolution of
the black hole outgoing radiation. These classical observers, experience
and describe physical phenomena in a classical language that needs
a time parameter. Hence, we need to consider a quantum black hole
in a semiclassical universe where time appears as an emergent quantity.

Time is the product of an approximation which aims to extract, from
the Wheeler-DeWitt equation, an external, semiclassical stage, in
which black hole and Hawking radiation quantum states evolve. 
In reference \cite{Kiefer2007} (section 5.4) we can find a derivation, from equations
(\ref{Wheeler-De Witt Eq})-(\ref{Diffeo quantum Eq}), of a Schrödinger functional equation.
In the following, we highlight some important details of this derivation.
Let us start by writing the wave functional as
\begin{equation}
|\Psi\left(h_{ab}\right)\rangle=e^{im_{\mathrm{Pl}}^2 S[h_{ab}]}|\psi\left(h_{ab}\right)\rangle
\: ,
\end{equation}
where $S[h_{ab}]$ is a solution of the vacuum Einstein-Hamilton-Jacobi function \cite{Barvinski1998}, since its WKB approximation enable us to extract, at higher orders, a Hamilton-Jacobi equation. In addition, $S[h_{ab}]$ is also solution to the Hamilton-Jacobi version of (\ref{Wheeler-De Witt Eq})-(\ref{Diffeo quantum Eq}), namely
\begin{eqnarray}
\dfrac{m_{\mathrm{Pl}}^2}{2}G_{abcd}\dfrac{\delta S}{\delta h_{ab}}\dfrac{\delta S}{\delta h_{cd}}
-2m_{\mathrm{Pl}}^2 \sqrt{h}\:\:{}^{\left(3\right)}R
+ \langle \psi| \hat{\mathcal{H}}{}_{\perp}^{\mathrm{m}} |\psi\rangle & = & 0 \: ,\\
-2m_{\mathrm{Pl}}^2 h_{ab} D_{c}\dfrac{\delta S}{\delta h_{bc}}
+ \langle \psi| \hat{\mathcal{H}}{}_{a}^{\mathrm{m}} |\psi\rangle & = & 0 \: ,
\end{eqnarray}
with the definitions $m_{\mathrm{Pl}}^2=(32\pi G)^{-1}$, $\hbar=c=1$ and $\hat{\mathcal{H}}{}_{\perp}^{\mathrm{m}}$ and $\hat{\mathcal{H}}{}_{a}^{\mathrm{m}}$ are assumed to be contributions from the non-gravitational fields. Having the solution $S[h_{ab}]$, we can now evaluate $|\psi\left(h_{ab}\right)\rangle$ along a solution of the classical Einstein equations $h_{ab}(\mathbf{x},t)$. In fact this solution is obtained from
\begin{equation}
\dot{h}_{ab}=N G_{abcd}\dfrac{\delta S}{\delta h_{cd}} + 2D_{(a}N_{b)} \: ,
\end{equation}
after a choice of the lapse and shift function has been made. At this point, we can define the evolutionary equation for the quantum state $|\psi\left(h_{ab}\right)\rangle$ as
\begin{equation}
\dfrac{\partial}{\partial t} |\psi\left(h_{ab}\right)\rangle = 
\int d^{3}x \: \dot{h}_{ab} \dfrac{\delta}{\delta h_{ab}}|\psi\left(h_{ab}\right)\rangle \: ,
\label{evol-wavefunct}
\end{equation} 
which, since $\dot{h}_{ab}$ depends on the DeWitt metric $G_{abcd}$, will have differential operators with the wrong sign in its right hand side. Finally, we are in the position of defining a functional Schrödinger equation for quantized matter fields in an external classical gravitational field as
\begin{eqnarray}
i\hbar\frac{\partial}{\partial t}\,|\psi\left[h_{ab}(\mathbf{x},t) \right]\rangle & = & \hat{H}{}^{\mathrm{m}}|\psi\left[h_{ab}(\mathbf{x},t) \right]\rangle\ ,\nonumber \\
\hat{H}{}^{\mathrm{m}} & \equiv & \int d^{3}x\left\{ N(\mathbf{x})\hat{\mathcal{H}}{}_{\perp}^{\mathrm{m}}(\mathbf{x})+N^{a}(\mathbf{x})\hat{\mathcal{H}}{}_{a}^{\mathrm{m}}(\mathbf{x})\right\} \ .\label{semi Schr equation}
\end{eqnarray}
Notice that the matter hamiltonian $\hat{H}{}^{\mathrm{m}}$, is parametrically depending on metric coefficients of the curved space-time background and contain indefinite kinetic terms. 

This derivation assumes a separation of the complete system (which state obeys the Wheeler-DeWitt equation and the quantum diffeomorphism invariance) in two parts,
in total correspondence with the way a Born-Oppenheimer approximation
is implemented. The physical system separation into two parts, one
purely quantum and the other semiclassical, is essentially achieved
by separating the gravitational from the non gravitational degrees
of freedom through an expansion, with respect to the Planck mass $m_{\mathrm{Pl}}$,
of constraints (\ref{Wheeler-De Witt Eq})-(\ref{Diffeo quantum Eq}).
However, we notice that there are gravitational degrees of freedom
that can be included in the purely quantum part (quantum density fluctuations
whose origin is gravitational, for example). Equation (\ref{semi Schr equation}),
formally similar to Schrödinger equation, is an equation with functional
derivatives, in which variable $\mathbf{x}$ is related to the 3-dimensional
metric $h_{ab}$. As previously mentioned, we recall that due to the
DeWitt $G_{abcd}$ metric definition, a negative kinetic term emerges
from the conjugated momentum $p_{ab}$.

In the following section, let us develop a simple model of the black
hole evaporation stage \cite{Kiefer:2008av}, 
which incorporates one interesting feature of the Wheeler-DeWitt equation, namely the indefinite kinetic term, and study some of its consequences. The main objective, here, is to estimate the degree of entanglement between Hawking radiation and
black hole quantum states, when we take into account a simple form
of back reaction between the two.

\section{Simplified model with a Schrödinger type of equation}
\label{section4}
Equation (\ref{semi Schr equation}) is a functional  differential equation, its wave functional solution depends on the 3-metric $h_{ab}$ describing the black hole and matter fields. It is obviously an almost impossible task to solve and find solutions to that equation. However, we can consider a simpler model, assuming a Schrödinger type of equation, which was first considered in \cite{Kiefer2007}. In that work it was argued that in order to study the effect of the indefinite kinetic term in (\ref{semi Schr equation}), as a first approach, and since we are dealing with an equation which is formally a Schrödinger equation, we could restrict our attention to finite amount of degrees of freedom. This first approach as been successful in cosmology, allowing to solve the Wheeler-DeWitt equation in minisuperspace, which brings a functional differential equation to a regular differential equation. We do not claim that we are doing the exact same process, but instead that a reduction of the physical system to a finite number of degrees of freedom could retain some aspects of quantum gravity that could be studied using much simpler equations. It is an acceptable concern if approximating a functional differential equation to a Schrödinger type of regular differential equation becomes an oversimplification. Nevertheless, it can also be acceptable to think that some physical insight can be obtained by assuming that the indefinite character of the functional equation is mimicked in the simpler model. Let us consider some assumptions in order to obtain the simpler equation. 

\begin{figure}[h]
\centering{}\includegraphics[width=6cm]{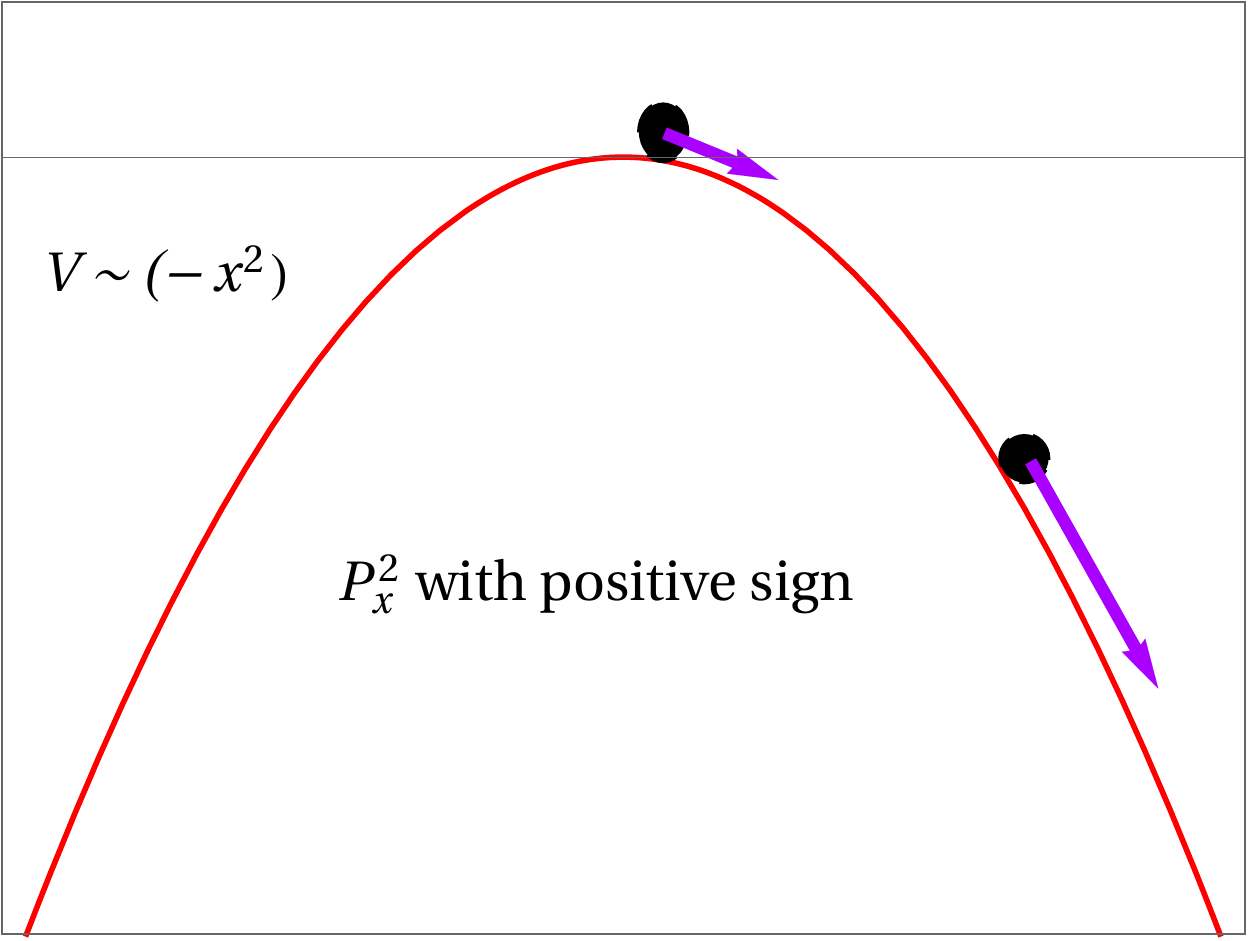}\quad{}\quad{}\includegraphics[width=6cm]{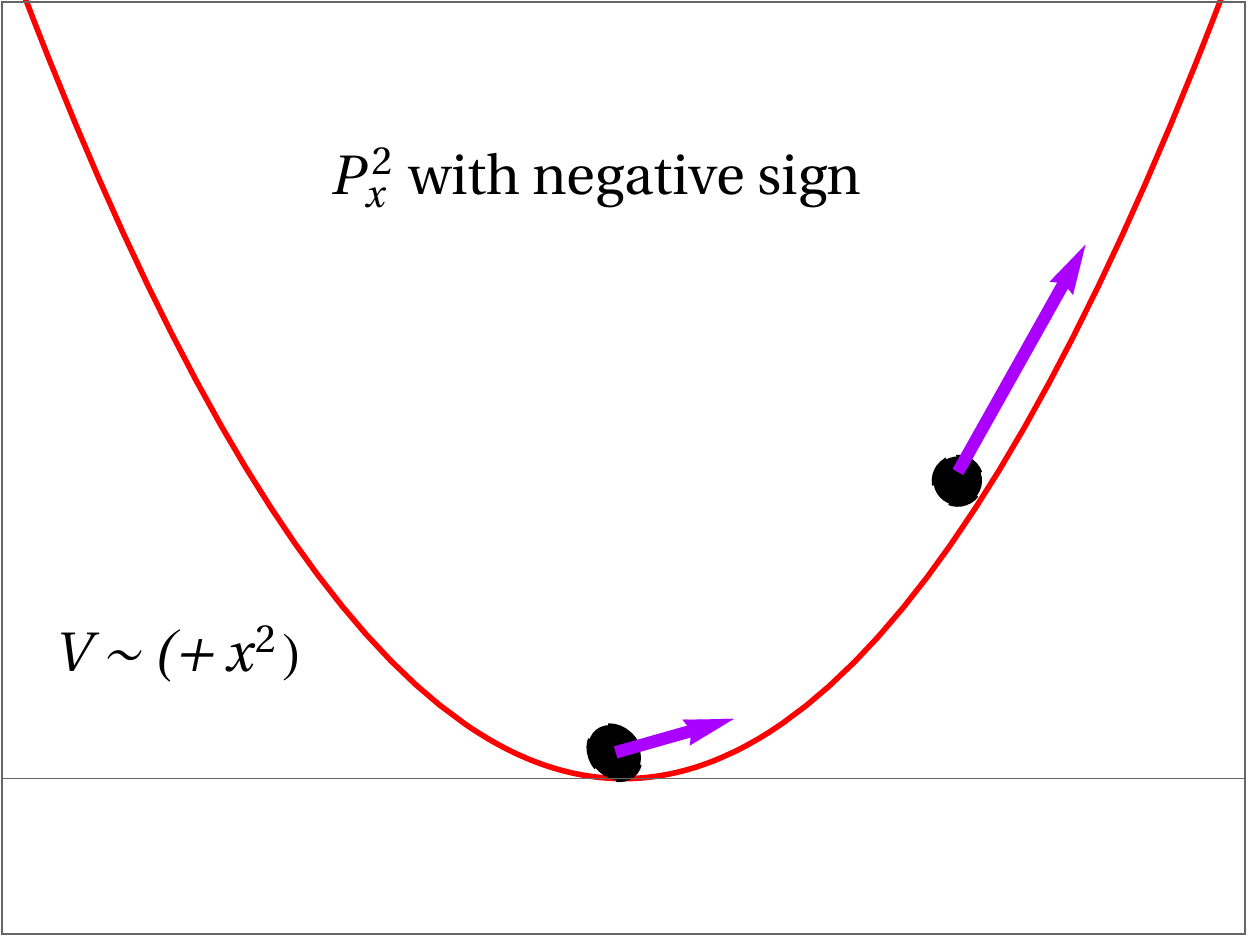}
\caption{{\footnotesize{}\label{Inverted oscillator} The behaviour of a particle
in an inverted oscillator potential, with a positive kinetic term,
is equivalent to the behaviour of a particle, with a negative kinetic
term, in a regular harmonic oscillator potential.}}
\end{figure}

\begin{enumerate}
\item Assuming that the hamiltonian $\hat{H}{}^{\mathrm{m}}$ includes black
hole and Hawking radiation parts, and ignoring other degrees of freedom,
the simpler equation can take the form,
\begin{eqnarray}
i\hbar\frac{\partial}{\partial t}\Psi(x,y,t) & = & \left(\frac{\hbar^{2}}{2m_{\mathrm{Pl}}}\frac{\partial^{2}}{\partial x^{2}}-\frac{\hbar^{2}}{2m_{y}}\frac{\partial^{2}}{\partial y^{2}}+\frac{m_{\mathrm{Pl}}\omega_{x}^{2}}{2}x^{2}+\frac{m_{y}\omega_{y}^{2}}{2}y^{2}\right)\Psi(x,y,t)\ .\label{model}
\end{eqnarray}
This last equation, where the emergence of a negative kinetic term
which plays the role of the functional derivative in the metric $h_{ab}$
in (\ref{Wheeler-De Witt Eq}), contrasts with an exact Schrödinger
equation. Therefore, because variable $x$ is related to metric $h_{ab}$,
we propose to identify it with the variation of the black hole radius\footnote{A black hole without rotation and charge which is simply described
by the Schwarzschild static solution.} $2GM/c^{2}$, which turn out to be also a variation in the black
hole mass or energy. Variable $y$ will correspond to Hawking radiation
with energy $m_{y}$.
\item Notice that the kinetic term of the gravitational part of the hamiltonian operator is suppressed by the Planck mass. As long as the black hole mass is large, this kinetic term is irrelevant. One would have in that case, only the Hawking radiation contribution. If, instead we consider the last stages of the evaporation process, when the black hole mass approaches the Planck mass, then the kinetic term associated with the black hole state becomes relevant. 
\item The time parameter $t$ in equation (\ref{model}) was obtained by
means of a Born-Oppenheimer approximation and embodies all the semiclassical
degrees of freedom of the universe.
\item In equation (\ref{model}) we consider harmonic oscillator potentials.
Beside being simpler potentials, they allow for analytical solutions
and, in the Hawking radiation case this regime is realistic \cite{Demers_1996,Kiefer_2001}.
For the black hole, this potential is an oversimplification, which
can be far from realistic. However, it can help to disclose behaviours
also present among more complex potentials, with respect to the entanglement
between black hole and Hawking radiation quantum states, during the
evaporation process. Furthermore, before dealing with the full problem,
simpler models can identify physical phenomena that will reasonably
manifest independently of the problem complexity (for example, the
infinite square well helps to understand energy quantisation in the
more complex Coulomb potential).
\end{enumerate}
Let us assume that equation (\ref{model}) is solved by the variables
separation method,
\begin{equation}
\Psi(x,y,t)=\psi_{x}(x,t)\psi_{y}(y,t),\label{Variable separation}
\end{equation}
so that we can obtain the two following equations\footnote{Where $\psi_{x}^{*}$ is the complex conjugate of $\psi_{x}$.},
\begin{equation}
\begin{aligned}i\hbar\dot{\psi}_{x}^{*}(x,t) & =\left(-\frac{\hbar^{2}}{2m_{\mathrm{Pl}}}\frac{\partial^{2}}{\partial x^{2}}-\frac{m_{\mathrm{Pl}}\omega_{x}^{2}}{2}x^{2}\right)\psi_{x}^{*}(x,t)\\
i\hbar\dot{\psi}_{y}(y,t) & =\left(-\frac{\hbar^{2}}{2m_{y}}\frac{\partial^{2}}{\partial y^{2}}+\frac{m_{y}\omega_{y}^{2}}{2}y^{2}\right)\psi_{y}(y,t)
\end{aligned}
\:.\label{Separate Schr. equations}
\end{equation}
Equations (\ref{Separate Schr. equations}) describe an uncoupled
system comprising a harmonic oscillator and an inverted one. In figure
\ref{Inverted oscillator} 
 we illustrate the fact that having regular harmonic potential with
a negative (indefinite) kinetic term is equivalent, in the quantum
point of view, to the situation where an inverted oscillator potential
has a positive kinetic term. In both situations we have to deal with
an unstable system, which would correspond of having variable $x$
varying uncontrollably. A wave function $\psi_{0}(x^{\prime},0)$
that initially has a Gaussian profile, will evolve over time according
to,

\begin{equation}
\int dx^{\prime}\ G_{\mathrm{inv.}}(x,\,x^{\prime};\,t,\,0)\,\psi_{0}(x^{\prime},\,0)=\psi(x,\,t)\ ,\label{Inverted osc propagation}
\end{equation}
where $G_{\mathrm{inv.}}(x,\,x^{\prime};\,t,\,0)$ is the inverted
oscillator Green function \cite{Muller-Kirsten_2012,Kramer_2005},
\begin{equation}
G_{\mathrm{inv.}}(x,x^{\prime};t,0)=\sqrt{\frac{m_{\mathrm{Pl}}\omega_{x}}{2\pi i\hbar\sinh(\omega_{x}t)}}\exp\left(im_{\mathrm{Pl}}\omega_{x}\frac{\left((x^{2}+x^{\prime}{}^{2})\cosh(\omega_{x}t)-2xx^{\prime}\right)}{2\hbar\sinh(\omega_{x}t)}\right)\ ,\label{Inverted osc propagator}
\end{equation}
which can be obtained from the harmonic oscillator Green function
\begin{equation}
G_{\mathrm{osc.}}(x,x^{\prime};t,0)=\sqrt{\frac{m_{y}\omega_{y}}{2\pi i\hbar\sin(\omega_{y}t)}}\exp\left(im_{y}\omega_{y}\frac{\left((x^{2}+x^{\prime}{}^{2})\cos(\omega_{y}t)-2xx^{\prime}\right)}{2\hbar\sin(\omega_{y}t)}\right)\ ,\label{Regular osc propagator}
\end{equation}
by redefining $\omega\rightarrow\left(i\omega\right)$. The wave function
obtained from the computation of equation (\ref{Inverted osc propagation})
shows a progressive squeezing of the state in phase space, which means
an increasing uncertainty in the value of $x$. Physically, in this
simplified model, that would correspond to an unstable variation of
the Schwarzschild radius or mass of the black hole. Even though, conceivably,
a strong squeezing of the black hole state would occur \cite{Grishchuk1990},
driving its disappearance.

In the next section we will introduce the effect of a back reaction,
coupling effectively black hole and Hawking radiation quantum states,
and see that, under particular circumstances, the system becomes stable
and strongly entangled.

\section{Back reaction and Schrödinger equation}
\label{section5}
In this section we review and reproduce some results obtained in reference \cite{Kiefer:2008av}. Notice that in this work a slight change in some definitions will be carried. In addition some new aspects of the model will be discussed. 
In order to investigate the effects of a back reaction between Hawking
radiation and black hole states, let us consider a linear coupling
$\mu xy$ between the variables, where $\mu$ is a constant, in equation

\begin{eqnarray}
i\hbar\frac{\partial}{\partial t}\Psi(x,y,t) & = & \left(\dfrac{P_{y}^{2}}{2m_{y}}-\dfrac{P_{x}^{2}}{2m_{\mathrm{Pl}}}+\frac{m_{\mathrm{Pl}}\omega_{x}^{2}}{2}x^{2}+\frac{m_{y}\omega_{y}^{2}}{2}y^{2}+\mu xy\right)\Psi(x,y,t)\:.\label{model back react}
\end{eqnarray}
We should emphasize that, following the Born-Oppenheimer and WKB approximation used to obtain Eq. (\ref{semi Schr equation}), any phenomenological back reaction effect, here parametrized by $\mu$ must be suppressed by the Planck mass \citep{Kiefer2007}. Therefore we can consider that this back reaction coupling constant, as the kinetic term of the gravitational part of the hamiltonian operator, only becomes relevant when the black hole approaches the Planck mass. Consequently we can assume that the constant $\mu \thicksim \mu'/m_{\mathrm{Pl}}$.  
Suppose the initial state, describing the black hole, is the coherent
state 
\begin{equation}
\psi_{x_{0}}^{\alpha}(x,\,0)=\left(\frac{m_{\mathrm{Pl}}\omega_{x}}{\pi\hbar}\right)^{1/4}\exp\left(-\frac{m_{\mathrm{Pl}}\omega_{x}}{2\hbar}x^{2}+\alpha\sqrt{\frac{2m_{\mathrm{Pl}}\omega_{x}}{\hbar}}x-\frac{\vert\alpha\vert^{2}}{2}-\frac{\alpha^{2}}{2}\right)\:,\label{BH coherent state}
\end{equation}
where 
\begin{equation}
\alpha=\sqrt{\frac{m_{\mathrm{Pl}}\omega_{x}}{2\hbar}}x_{0}+i\frac{p_{0}}{\sqrt{2m_{\mathrm{Pl}}\omega_{x}}}\:,
\end{equation}
which represents a black hole whose Schwarzschild radius oscillates
around the value $2GM/c^{2}$. A coherent state represents a displacement
of the harmonic oscillator ground state $\left|0\right\rangle $ 
\begin{equation}
\left|\alpha\right\rangle =\hat{D}\left(\alpha\right)\left|0\right\rangle =e^{-\alpha\hat{a}^{\dagger}-\alpha^{*}\hat{a}}\left|0\right\rangle \:,\label{Desloc estado coerente}
\end{equation}
in order to get a finite excitation amplitude $\alpha$. For the Hawking
radiation initial state, let us consider the Gaussian distribution
\begin{equation}
\psi_{y_{0}}^{H}(y,\,0)\propto\exp\left(-\frac{m_{y}\omega_{y}}{2\hbar}\coth\left(\frac{2\pi\omega_{y}\mathrm{GM}}{c^{3}}\right)y^{2}\right)\:,\label{Hr state}
\end{equation}
which describes the radiation state \cite{Grishchuk1990,Demers_1996,Kiefer_2001}
for a black hole with Schwarzschild radius $2GM/c^{2}$. Under these
conditions, we can expect that, after the product state (\ref{BH coherent state})-(\ref{Hr state})
evolves in time
\begin{equation}
\hat{\mathcal{U}}\left|\Psi_{0}\right\rangle =\hat{\mathcal{U}}\left(\left|\psi_{x_{0}}^{\alpha}\right\rangle \otimes\left|\psi_{y_{0}}^{H}\right\rangle \right)=\left|\Psi\right\rangle \:,
\end{equation}
the emerging final state $\left|\Psi\right\rangle$ will be entangled,
because the hamiltonian in equation (\ref{model back react}) includes
a coupling such that $\hat{\mathcal{H}}\neq\hat{\mathcal{H}}_{x} \otimes\mathbbm{1} + \mathbbm{1} \otimes \hat{\mathcal{H}}_{y}$.

Determining the initial state $\left|\psi_{x_{0}}^{\alpha}\right\rangle \otimes\left|\psi_{y_{0}}^{H}\right\rangle $
evolution over time would be solved if we had the propagator related
to the hamiltonian of equation (\ref{model back react}). Since this
propagator is not available, we can instead redefine variables 
\begin{equation}
\left(\begin{array}{c}
P_{x}\\
P_{y}\\
x\\
y
\end{array}\right)=\left(\begin{array}{cccc}
\sqrt{\frac{2m_{\mathrm{Pl}}}{\cos2\theta}}\cos\theta & \sqrt{\frac{2m_{\mathrm{Pl}}}{\cos2\theta}}\sin\theta & 0 & 0\\
\sqrt{\frac{2m_{y}}{\cos2\theta}}\sin\theta & \sqrt{\frac{2m_{y}}{\cos2\theta}}\cos\theta & 0 & 0\\
0 & 0 & \frac{\cos\theta}{\sqrt{m_{\mathrm{Pl}}}\cos2\theta} & \frac{\sin\theta}{\sqrt{m_{\mathrm{Pl}}}\cos2\theta}\\
0 & 0 & \frac{\sin\theta}{\sqrt{m_{y}}\cos2\theta} & \frac{\cos\theta}{\sqrt{m_{y}}\cos2\theta}
\end{array}\right)\:\left(\begin{array}{c}
P_{1}\\
P_{2}\\
Q_{1}\\
Q_{2}
\end{array}\right)\:,\label{Variable redefinition}
\end{equation}
such that we can rewrite equation (\ref{model back react}) in the
following way
\begin{equation}
i\hbar\frac{\partial}{\partial t}\Psi(Q_{1},\:Q_{2},\:t)=\left(\frac{1}{2}\left(P_{2}^{2}-P_{1}^{2}\right)+\frac{1}{2}\left(\Omega_{1}^{2}Q_{1}^{2}+\Omega_{2}^{2}Q_{2}^{2}\right)+\mathcal{K}Q_{1}Q_{2}\right)\Psi(Q_{1},\:Q_{2},\:t)\:.\label{Schr in Q variable}
\end{equation}
In the previous equation, coordinates redefinition (\ref{Variable redefinition})
implies that
\begin{equation}
\begin{aligned}\Omega_{1}^{2}\cos^{2}2\theta & =\omega_{x}^{2}\cos^{2}\theta+\omega_{y}^{2}\sin^{2}\theta+\frac{\mu\sin2\theta}{\sqrt{m_{\mathrm{Pl}}m_{y}}}\\
\Omega_{2}^{2}\cos^{2}2\theta & =\omega_{x}^{2}\sin^{2}\theta+\omega_{y}^{2}\cos^{2}\theta+\frac{\mu\sin2\theta}{\sqrt{m_{\mathrm{Pl}}m_{y}}}
\end{aligned}
\:,\label{New frequency}
\end{equation}
and the coupling is
\begin{equation}
\mathcal{K}=\frac{1}{\cos^{2}2\theta}\biggl(\frac{1}{2}\left(\omega{}_{x}^{2}+\omega_{y}^{2}\right)\sin2\theta+\frac{\mu}{\sqrt{m_{\mathrm{Pl}}m_{y}}}\biggr)\:.\label{Q coupling}
\end{equation}
If we impose that, in the new variables $\left(Q_{1},\,Q_{2}\right)$,
the coupling is $\mathcal{K}=0$, it follows that the coupling in
the original variables $\left(x,\,y\right)$ is given by,
\begin{equation}
\mu=-\frac{1}{2}\sqrt{m_{\mathrm{Pl}}m_{y}}\left(\omega{}_{x}^{2}+\omega_{y}^{2}\right)\sin2\theta\:,\label{XY coupling}
\end{equation}
with $\theta\in]-\frac{\pi}{4},\frac{\pi}{4}[$. We can check that
$\mu=0$ for $\theta=0$. In the numerical simulations, to calculate
the relevant physical quantities, we will assume that $m_{y}=10^{-5}m_{\mathrm{Pl}}$
and $\omega_{y}^{2}=10^{5}\omega{}_{x}^{2}$ such that the potentials,
in equation (\ref{model back react}), are of the same order, i.e.,
$m_{y}\omega_{y}^{2}\sim m_{\mathrm{Pl}}\omega{}_{x}^{2}$. This corresponds
to assume that the Hawking radiation energy is well below Planck scale
and, the fluctuations of the Schwarzschild radius have significantly
smaller frequency than the Hawking radiation energy fluctuations.
The numerical factor choice of $10^{5}$ is arbitrary and does not
influence the conclusions to be drawn from the results presented in
subsequent sections. However, we can establish that the coupling is
defined in the interval 
\begin{equation}
-10^{2}\leq\mu\leq10^{2}\:,
\end{equation}
which is sufficiently broad to explore the more relevant cases. If
we substitute the coupling equation (\ref{XY coupling}) in the frequencies
definition (\ref{New frequency}), we will obtain, in the new coordinates,
\begin{align}
\Omega_{1}^{2} & =\frac{1}{\cos^{2}2\theta}\biggl[\omega_{x}^{2}\biggl(\cos^{2}\theta-\frac{1}{2}\sin^{2}2\theta\biggr)+\omega_{y}^{2}\biggl(\sin^{2}\theta-\frac{1}{2}\sin^{2}2\theta\biggr)\biggr]\ ,\nonumber \\
\Omega_{2}^{2} & =\frac{1}{\cos^{2}2\theta}\biggl[\omega_{x}^{2}\biggl(\sin^{2}\theta-\frac{1}{2}\sin^{2}2\theta\biggr)+\omega_{y}^{2}\biggl(\cos^{2}\theta-\frac{1}{2}\sin^{2}2\theta\biggr)\biggr].\label{New frequency - v1}
\end{align}
We can notice an important observation related to equation (\ref{New frequency - v1}).
Since, we assume that $\omega_{y}^{2}\gg\omega_{x}^{2}$, it implies
that $\Omega_{1}^{2}$ remains strictly positive\footnote{
Whereas $\Omega_{2}^{2}$ is always positive, since the definition of $\Omega_2^2$ can be further simplified to \\
$\Omega_2^2=\dfrac{1}{2}\left( 1-\dfrac{1}{\cos 2\theta}\right)+\dfrac{m_{\mathrm{Pl}}}{2m_y}\left( 1+\dfrac{1}{\cos 2\theta}\right)\thicksim \dfrac{10^5}{2}\left( 1+\dfrac{1}{\cos 2\theta}\right)$\\
which for $\theta\in]-\frac{\pi}{4},\frac{\pi}{4}[$ is always positive.
} in the significantly reduced sub interval of the possible angles $\theta\in]-\frac{\pi}{4},\frac{\pi}{4}[$.
We can verify that $\Omega_{1}^{2}$ is only positive when
\begin{equation}
-\arctan\left(\sqrt{\dfrac{\omega_{x}^{2}}{\omega_{y}^{2}}}\right)<\theta<\arctan\left(\sqrt{\dfrac{\omega_{x}^{2}}{\omega_{y}^{2}}}\right)\:.\label{Interval theta}
\end{equation}
This observation means that, for values outside the mentioned interval
(\ref{Interval theta}), $\Omega_{1}^{2}$ is negative and equation
(\ref{Schr in Q variable}) turns out to be a Schrödinger equation
describing two uncoupled harmonic oscillators in the coordinates $\left(Q_{1},\,Q_{2}\right)$,
i.e.
\begin{equation}
i\hbar\frac{\partial}{\partial t}\Psi(Q_{1},\:Q_{2},\:t)=\left(\frac{1}{2}\left(P_{2}^{2}+\Omega_{2}^{2}Q_{2}^{2}\right)-\frac{1}{2}\left(P_{1}^{2}+\left|\Omega_{1}^{2}\right|Q_{1}^{2}\right)\right)\Psi(Q_{1},\:Q_{2},\:t)\:.
\end{equation}
In addition, we also have 
\begin{equation}
\arctan\left(\sqrt{\dfrac{\omega_{x}^{2}}{\omega_{y}^{2}}}\right)<\left|\theta\right|<\dfrac{\pi}{4}\quad\Rightarrow\quad\left|\mu\right|>1
\end{equation}
which implies that, in equation (\ref{model back react}), when the
coupling is $\left|\mu\right|>1$ the system becomes stable and this
will restraint the influence of the inverted potential.

The calculation of the initial state $\left|\psi_{x_{0}}^{\alpha}\right\rangle \otimes\left|\psi_{y_{0}}^{H}\right\rangle $
time evolution, in coordinates $\left(Q_{1},\,Q_{2}\right)$, with
the help of the harmonic (\ref{Regular osc propagator}) and inverted
oscillator propagators,
\begin{equation}
\int\int dQ_{1}'dQ_{2}'\quad G_{\mathrm{inv.}}(Q_{1},Q_{1}';t)\cdot G_{\mathrm{osc.}}(Q_{2},Q_{2}';t)\cdot\psi_{x_{0}}^{\alpha}(Q_{1}',\,Q_{2}';\,0)\cdot\psi_{y_{0}}^{H}(Q_{1}',\,Q_{2}';\,0)\:,\label{Psi Q evolution}
\end{equation}
enable us to obtain $\Psi(Q_{1},\:Q_{2},\:t)$, for which an explicit
analytical expression is given in appendix \ref{apendice 2} (equation
(\ref{WaveFunv Q})). Subsequently, we can use the inverse transformation
\begin{equation}
\begin{cases}
Q_{1}= & x\sqrt{m_{\mathrm{Pl}}}\cos\theta-y\sqrt{m_{y}}\sin\theta\\
Q_{2}= & -x\sqrt{m_{\mathrm{Pl}}}\sin\theta+y\sqrt{m_{y}}\cos\theta
\end{cases}\:,\label{Q to X coord transform}
\end{equation}
in order to retrieve the wave function in the original coordinates.
This wave function has the generic form
\begin{equation}
\Psi(x,\,y,\,t)=F(t)\,\exp\left(-A(t)x^{2}+B(t)x-C(t)y^{2}+D\left(t\right)y+E(t)xy\right)\:,\label{Psi X evolution}
\end{equation}
where the time dependent functions can be found in appendix \ref{apendice 2},
more precisely in equation (\ref{TimeFunc2}). 

One of the main objectives in this paper is to quantify the entanglement
degree between black hole and Hawking radiation quantum states. In
order to proceed with that idea we have to define the system matrix
density
\begin{equation}
\rho_{xy}=\left|\Psi\right\rangle \left\langle \Psi\right|\:.\label{Density matrix}
\end{equation}
Wave function (\ref{Psi X evolution}) cannot be factored, hence,
the initial density matrix $\left|\Psi_{0}\right\rangle \left\langle \Psi_{0}\right|$,
corresponding to the factored pure state $\left|\psi_{x_{0}}^{\alpha}\right\rangle \otimes\left|\psi_{y_{0}}^{H}\right\rangle $,
has evolved to a pure entangled state described by $\rho_{xy}$. Recalling
the status of the classical observers outside the black hole, they
can only access the state of the outgoing radiation, i.e., they can
only experiment part of the system. Therefore, it is important to
consider the reduced density matrix $\rho_{\mathrm{y}}$ obtained
by taking the partial trace of the system density matrix $\rho_{xy}$,
i.e., computing $\rho_{\mathrm{y}}=\mathrm{tr}_{x}\,\left(\rho_{xy}\right)$.
The reduced density matrix elements, for black hole and Hawking radiation,
are respectively
\begin{equation}
\begin{aligned}\rho_{Bh}\left(x,\,x'\right)= & \mathrm{tr}_{y}\rho_{xy}=\int dy\:\left|\langle x',\,y|x,\,y\rangle\right|{}^{2}\\
\rho_{Hr}\left(y,\,y'\right)= & \mathrm{tr}_{x}\rho_{xy}=\int dx\:\left|\langle x,\,y'|x,\,y\rangle\right|{}^{2}
\end{aligned}
\:,
\end{equation}
where $|x,y\rangle\equiv\Psi(x,\,y,\,t)$, and with the generic form
\begin{equation}
\begin{aligned}\rho_{Bh}\left(x,\,x'\right)= & \mathcal{N}_{1}\exp\left(-\mathcal{A}_{1}x^{2}+\mathcal{B}_{1}x-\mathcal{A}_{1}^{*}x'^{2}+\mathcal{B}_{1}^{*}x'+\left|\mathcal{C}_{1}\right|xx'\right)\\
\rho_{Hr}\left(y,\,y'\right)= & \mathcal{N}_{2}\exp\left(-\mathcal{A}_{2}y^{2}+\mathcal{B}_{2}y-\mathcal{A}_{2}^{*}y'^{2}+\mathcal{B}_{2}^{*}y'+\left|\mathcal{C}_{2}\right|yy'\right)
\end{aligned}
\:.\label{Bh-Hr density matrix elements}
\end{equation}
The coefficients defined in the last equation are given in appendix
\ref{apendice 3} (equations (\ref{Bh Wigner coef})-(\ref{Hr Wigner coef})),
and also depend directly on equation (\ref{TimeFunc2}). 

\begin{figure}[h]
\centering{}\includegraphics[width=7cm]{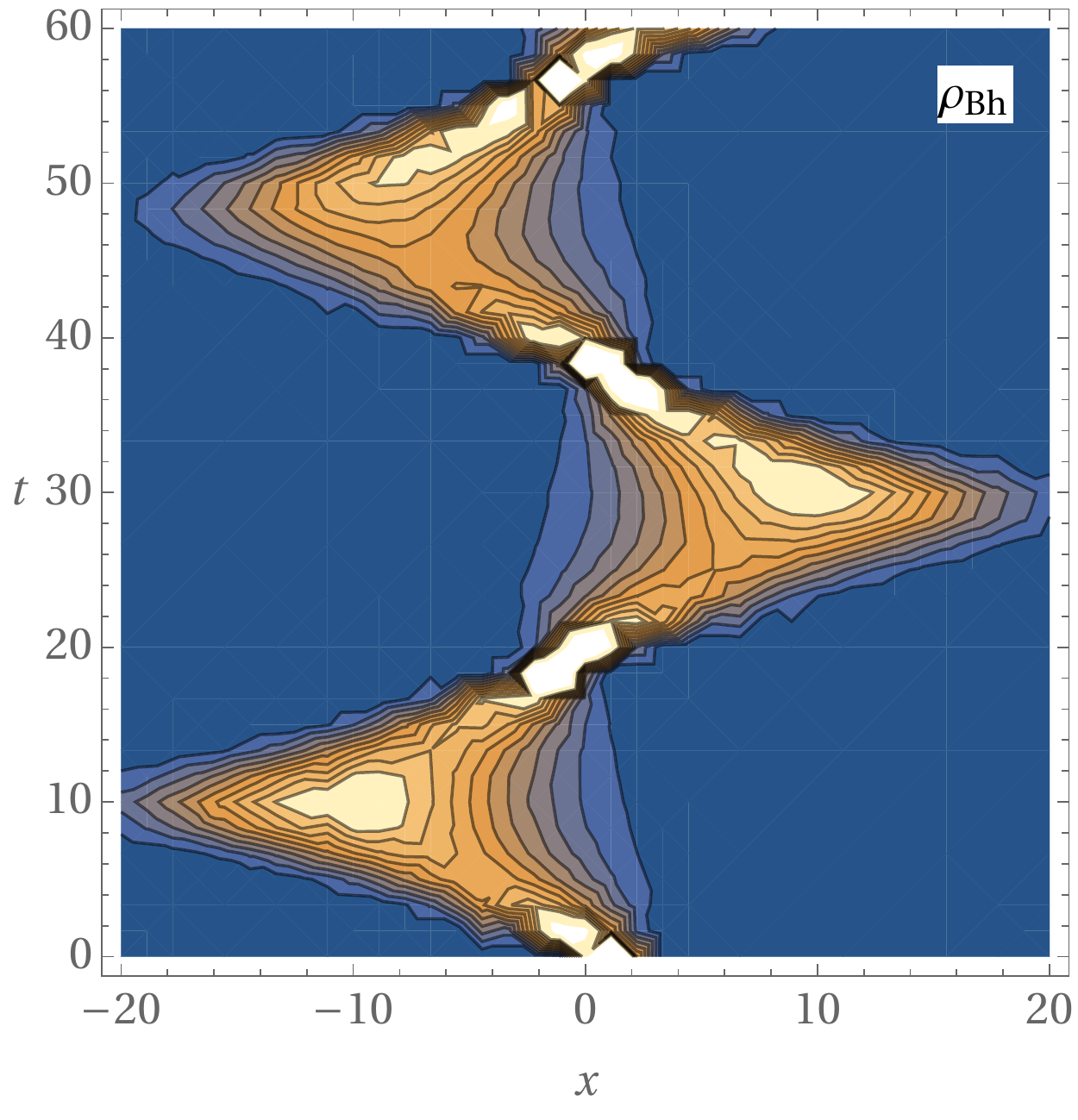}\quad{}\quad{}\includegraphics[width=7cm]{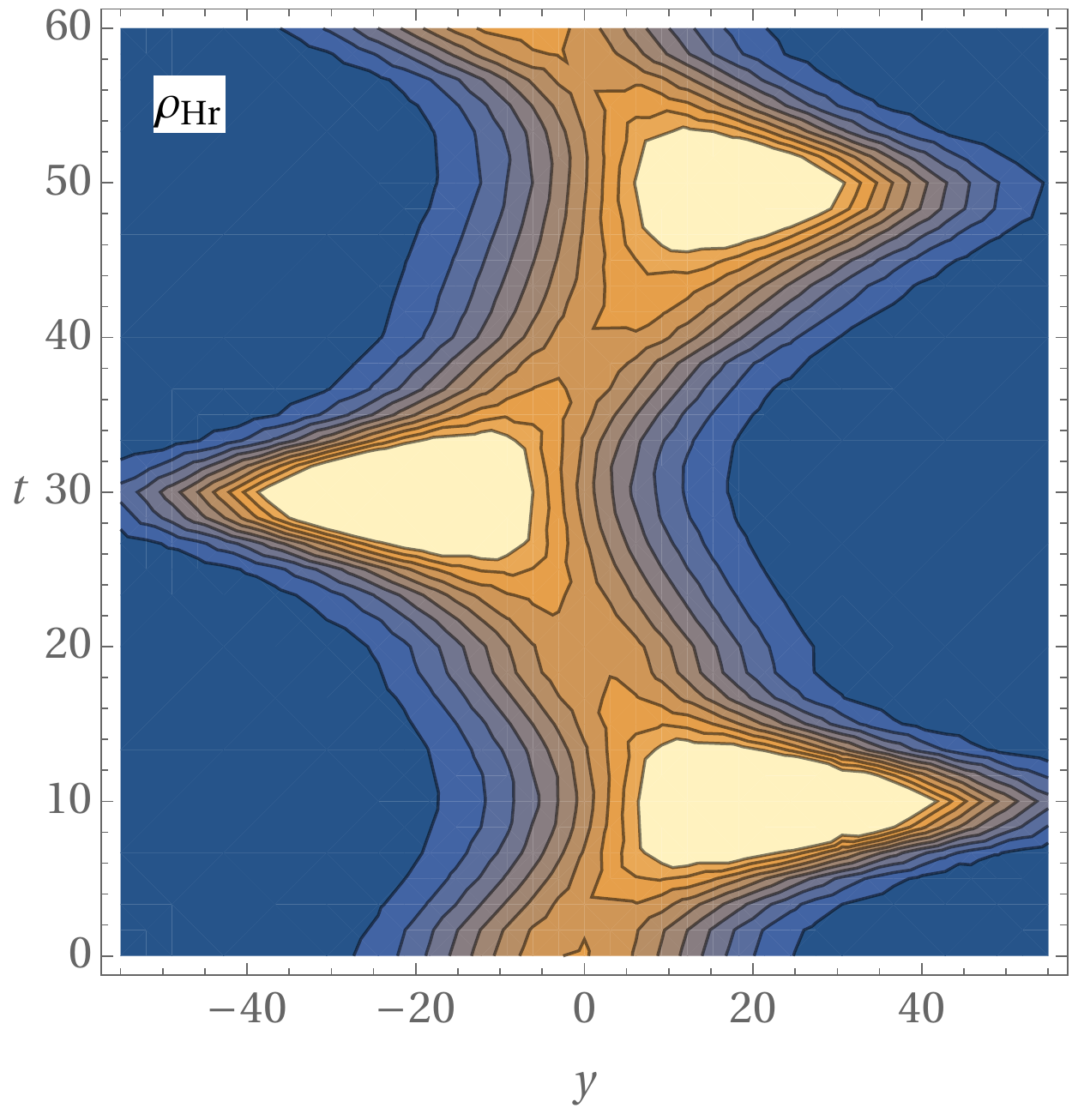}
\caption{{\footnotesize{}\label{Bh and Hr density - 1} Diagonal reduced density
matrix elements $\rho_{Bh}$ and $\rho_{Hr}$ (defined in equations
(\ref{Bh reduced density matrix diag})-(\ref{Hr reduced density matrix diag}))
evolution over time, with $m_{\mathrm{Pl}}=\hbar=\omega_{x}=x_{0}=1$;
$p_{0}=-1$. The back reaction coupling value is $\mu\approx1.01$,
where $\omega_{y}=\omega_{x}\times10^{5/2}$ and $m_{y}=m_{\mathrm{Pl}}\times10^{-5}$.
Light areas, in the density plots, correspond to higher values of
the density matrix.}}
\end{figure}

The diagonal reduced density matrix elements are
\begin{equation}
\rho_{Bh}\left(x,\,x\right)=|F|^{2}\sqrt{\frac{1}{2\mathrm{Re}\left(C\right)}}\exp\left(-2\mathrm{Re}\left(A\right)x^{2}+\frac{\left(\mathrm{Re}\left(E\right)x+\mathrm{Re}\left(D\right)\right)^{2}}{2\mathrm{Re}\left(C\right)}+2\mathrm{Re}\left(B\right)x\right)\label{Bh reduced density matrix diag}
\end{equation}
\begin{equation}
\rho_{Hr}\left(y,\,y\right)=|F|^{2}\sqrt{\frac{1}{2\mathrm{Re}\left(A\right)}}\exp\left(-2\mathrm{Re}\left(C\right)y^{2}+\frac{\left(\mathrm{Re}\left(E\right)y+\mathrm{Re}\left(B\right)\right)^{2}}{2\mathrm{Re}\left(A\right)}+2\mathrm{Re}\left(D\right)y\right)\label{Hr reduced density matrix diag}
\end{equation}
and, for illustration purposes, in figure \ref{Bh and Hr density - 1}
we can observe their evolution over time. In that case, we have taken
$\mu=1.01$ for the back reaction coupling value. As we emphasised
before, for this value, the system is stable and we can notice that
the observed behaviours corresponds closely to squeezed coherent states\footnote{This observation will be corroborated by inspecting the behaviour
of the Wigner functions in appendix \ref{apendice 3}. Squeezed coherent
states are obtained through the action of two different operators
over the ground state of the harmonic oscillator $\left|\alpha,\,\xi\right\rangle =\hat{D}\left(\alpha\right)\hat{S}\left(\xi\right)\left|0\right\rangle =\left(e^{-\alpha\hat{a}^{\dagger}-\alpha^{*}\hat{a}}\right)\left(e^{-\frac{1}{2}\xi^{*}\hat{a}^{2}-\frac{1}{2}\xi\hat{a}^{\dagger2}}\right)\left|0\right\rangle $,
where $\hat{D}\left(\alpha\right)$ is the displacement operator and
$\hat{S}\left(\xi\right)$ is the squeeze operator.}, with an evident correlation between them. 

\section{Entropy, entanglement and information}
\label{section6}
Theoretically, black holes emit radiation, when measured by an infinitely
distant observer, approximately with a black body spectrum with an
emission rate in a mode of frequency $\omega$
\begin{equation}
\Gamma\left(\omega\right)=\dfrac{\gamma\left(\omega\right)}{\exp\left(\dfrac{\hbar\omega}{k_{B}T_{H}}\right)\pm1}\dfrac{d^{3}k}{\left(2\pi\right)^{3}}\:,
\end{equation}
where the Hawking temperature is, 
\begin{equation}
T_{H}=\dfrac{\hbar c^{3}}{8\pi k_{B}GM}
\end{equation}
and the factor $\gamma\left(\omega\right)$ embodies the effect of
the non trivial geometry surrounding the black hole. Soon after this
discovery, D. N. Page made important numerical estimates \cite{Page:1976df,Page:1976ki,Page:1977um},
of various particle emission rates, for black holes with and without
rotation, and the evaporation average time for a black hole with mass
$M$. Later, he made important conjectures \cite{Page1993-Aug} about
the Von Neumann entropy
\begin{equation}
S_{VN}=-\mathrm{tr}\:\left(\rho\log\left(\rho\right)\right)\label{Von Neumann entropy}
\end{equation}
of a quantum subsystem described by the reduced density matrix $\rho_{A}=\mathrm{tr}_{B}\,\rho_{AB}$.
If the Hilbert space of a quantum system, in a pure initial random
state, has dimension $mn$, the average entropy of the subsystem of
smaller dimension $m<n$ is conjectured to be given by
\begin{equation}
S_{m,\,n}\simeq\log m-\dfrac{m}{2n}\:.\label{DPage conjecture}
\end{equation}
Therefore, the given subsystem will be near its maximum entropy $\log m$
whenever $m<n$. Afterwards, he applied this new conjecture to the
case of the black hole evaporation process \cite{Page1993-Dec}. Assuming
that initially Hawking radiation and black hole are in a pure quantum
state, described by the density matrix $\rho_{AB}$, he showed that
the Von Neumann entropies related to the reduced density matrices
(radiation - \emph{Hr} - and black hole - \emph{Bh }-),
\begin{equation}
S_{Hr}=-\mathrm{tr}\:\left(\rho_{Hr}\log\left(\rho_{Hr}\right)\right)\label{Hr quant. entropy}
\end{equation}
 
\begin{equation}
S_{Bh}=-\mathrm{tr}\:\left(\rho_{Bh}\log\left(\rho_{Bh}\right)\right)\label{Bh quant. entropy}
\end{equation}
display an information (defined as a measure of the departure of the
actual entropy from its maximum value),
\begin{equation}
I_{Hr}=\log m-S_{Hr}\quad I_{Bh}=\log n-S_{Bh}\:.\label{D. Page Information}
\end{equation}
\begin{figure}[h]
\centering{}\includegraphics[width=8cm]{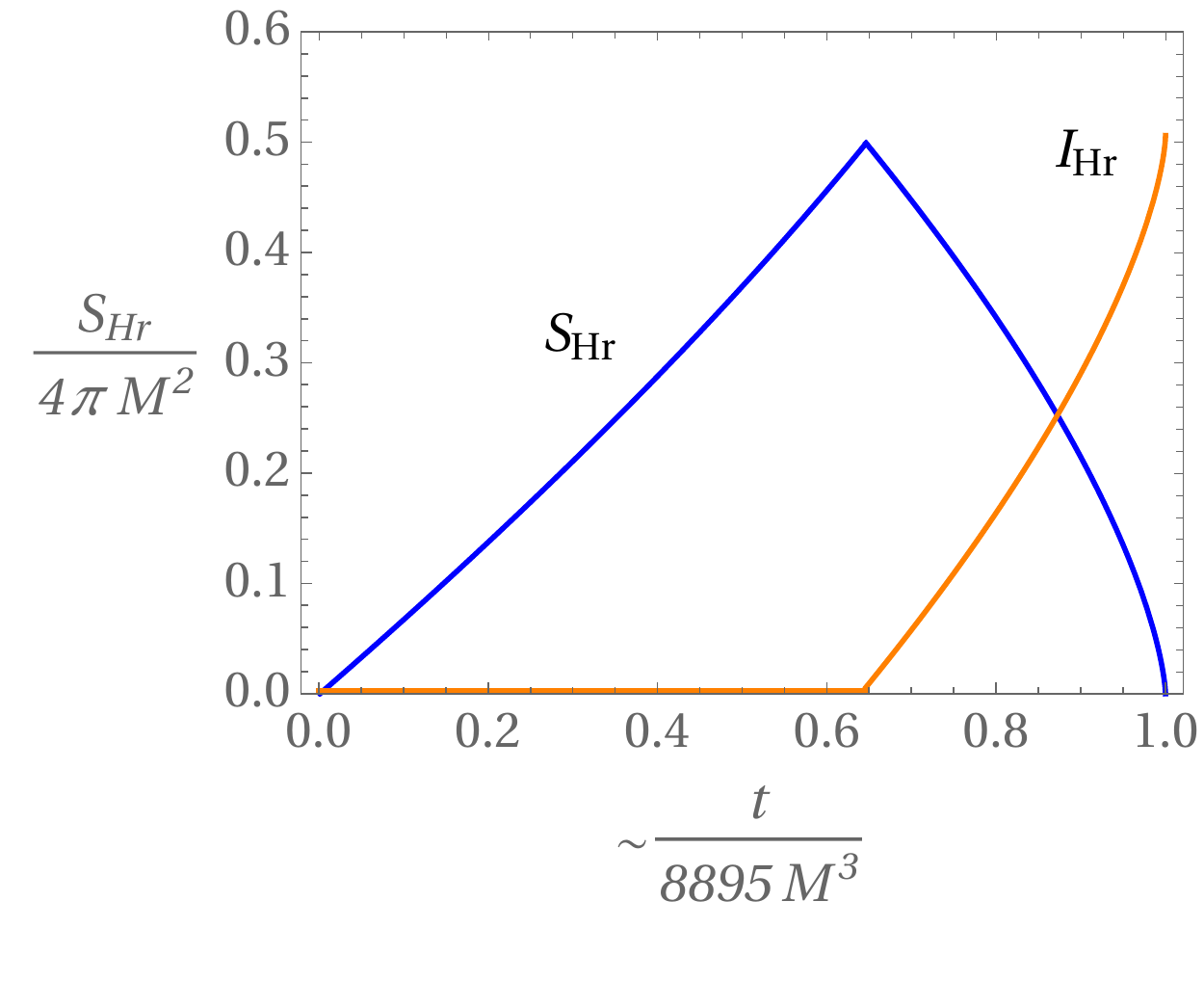}
\caption{{\footnotesize{}\label{Page curve} Time evolution of, the Von Neumann
entropy $\left(S_{Hr}\right)$ (according to Page curve \cite{Page1993-Dec,Page:2013dx}),
and the information $\left(I_{Hr}\right)$ for the Hawking radiation.}}
\end{figure}
In addition, he also described, through what is today known as the
Page curve (a recent nice review can be found in \cite{ALONSOSERRANO201810}),
the way entropy will evolve\footnote{Also, the way information included in the correlations between black
hole and Hawking radiation quantum states will evolve.} (see figure \ref{Page curve}) while the black hole evaporates. More
recently, he has numerically estimated, based on his previous works,
about emission rates of several types of particles, the way Hawking
radiation entropy should evolve in time \cite{Page:2013dx}. It is
believed that a correct quantum gravity theory should be able to show
how Page curve emerges from the assumption of the outgoing radiation
and black hole quantum states unitary evolution.

It seems pertinent to explore what the simplified model, under analysis,
allow us to say about entropy and information. More precisely, we
want to estimate how Hawking radiation entropy and information evolve
over time according to equation (\ref{D. Page Information}). Considering
the reduced density matrices (\ref{Bh-Hr density matrix elements}),
we see that to properly calculate Von Neumann entropy (\ref{Von Neumann entropy})
we have to diagonalize the matrices, i.e. compute their eigenvalues
\begin{equation}
\int_{-\infty}^{+\infty}dy'\,\rho_{Hr}\left(y,\,y'\right)f_{n}\left(y'\right)=\lambda_{n}f_{n}\left(y\right)\:.\label{Eigenvalue equation}
\end{equation}
This particular calculation is only known for a few specific cases,
as for example, for a system of two coupled harmonic oscillators \cite{Srednicki_1993},
unfortunately a distinct situation from the case studied here, namely
the coupling between harmonic and inverted oscillators. Solving the
eigenvalues problem allows a great simplification and the evaluation
of Von Neumann entropy becomes simply
\begin{equation}
S_{VN}=-\mathrm{\sum_{n}}\:\left(\lambda_{n}\log\left(\lambda_{n}\right)\right)\:.\label{VN entropy eigen}
\end{equation}
However, considering the eigenvalues problem technical difficulty,
instead of computing the Von Neumann entropy we can estimate the Wehrl
entropy \cite{Wehrl_1978,Wehrl_1979}, 
\begin{equation}
\begin{aligned}S_{W} & =-\mathrm{tr}\left(H_{Bh}\left(x,\,p\right)\log\left(H_{Bh}\left(x,\,p\right)\right)\right)\\
 & =\iint\dfrac{dxdp}{\pi\hbar}\:H_{Bh}\left(x,\,p\right)\log\left(H_{Bh}\left(x,\,p\right)\right)
\end{aligned}
\:,\label{Wehrl entropy}
\end{equation}
where $H_{Bh}\left(x,\,p\right)$ is the Husimi function \cite{Husimi}
\begin{equation}
\begin{aligned}H_{Bh}\left(x,\,p\right) & =\int\dfrac{dx'dp'}{\pi\hbar}\:\exp\left(-\dfrac{\sigma\left(x-x'\right)^{2}}{\hbar}-\dfrac{\left(p-p'\right)^{2}}{\sigma\hbar}\right)W_{Bh}\left(x',\,p'\right)\end{aligned}
\:.\label{Husimi function}
\end{equation}
The Husimi function is defined to access the classical phase space
$\left(x,\,p\right)$ representation of a quantum state, and it is
obtained from a Gaussian average of the Wigner function
\begin{equation}
\begin{aligned}W_{Bh}\left(x,\,p\right) & =\dfrac{1}{\pi\hbar}\int d\eta\:e^{-ip\eta/\hbar}\left\langle x+\dfrac{\eta}{2}\left|\rho_{Bh}\right|x-\dfrac{\eta}{2}\right\rangle \\
 & =\dfrac{1}{\pi\hbar}\int d\eta\:e^{-ip\eta/\hbar}\rho_{Bh}\left(x+\dfrac{\eta}{2},\,x-\dfrac{\eta}{2}\right)
\end{aligned}
\:.\label{Wigner function}
\end{equation}
The Wigner function give us an rough criterion on how much a quantum
state is distant from its classical limit but, unfortunately, it is
not a strictly positive function and cannot be taken as a probability
distribution in phase space\footnote{In fact, Wigner function is considered a quasiprobability distribution.}.
The Husimi representation (which is a Weierstrass transformation of
Wigner function) enable us to define a strictly positive function
and corresponds to the trace of the density matrix over the coherent
states basis $\left|\alpha\right\rangle $, i.e.,
\begin{equation}
H_{\alpha}=\dfrac{1}{\pi}\left\langle \alpha\left|\rho\right|\alpha\right\rangle \:.\label{Husimi coherent}
\end{equation}
If we compare this last definition with equations (\ref{VN entropy eigen})
and (\ref{Wehrl entropy}), we can understand Wehrl entropy as a classical
estimate of the Von Neumann entropy, through the analogy
\begin{equation}
S_{VN}=-\mathrm{\sum_{n}}\:\left(\lambda_{n}\log\left(\lambda_{n}\right)\right)\rightarrow S_{W}\approx-\mathrm{\sum}\:\left(H_{\alpha}\log\left(H_{\alpha}\right)\right)\:.
\end{equation}
Hence, Wehrl's entropy can be considered a measure of the classical
entropy of a quantum system, and has already been used \cite{Rosu:1994af}
in the contexts of cosmology and black holes. We should notice that
Wehrl entropy gives an upper bound to Von Neumann entropy, i.e., $S_{W}\left(\rho\right)\geq S_{VN}\left(\rho\right)$.
\begin{figure}[h]
\centering{}\includegraphics[width=8cm]{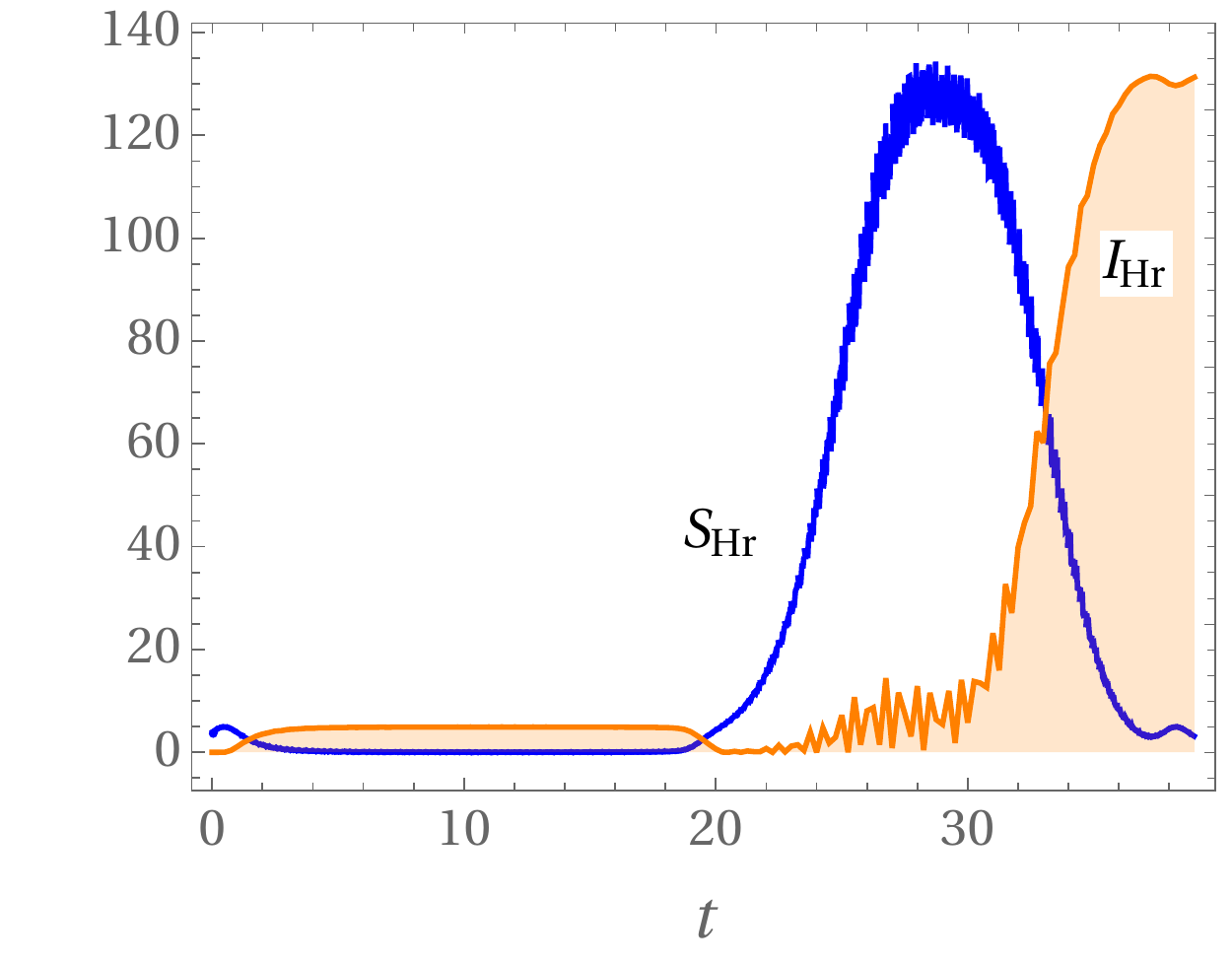}\quad{}\quad{}
\caption{{\footnotesize{}\label{ Hr entropy - 1} Hawking radiation's Wehrl
entropy and information evolution over time, with $m_{\mathrm{Pl}}=\hbar=\omega_{x}=x_{0}=1$;
$p_{0}=-1$. The back reaction coupling parameter is
$\mu=1.01$, where $\omega_{y}=\omega_{x}\times10^{5/2}$ and $m_{y}=m_{\mathrm{Pl}}\times10^{-5}$.}}
\end{figure}

The time has come to obtain, in this simplified model, Wehrl entropy
and information evolution over time for Hawking radiation. In figure
\ref{ Hr entropy - 1} we can find the numerical estimates of Hawking
radiation Wehrl entropy and information. These were obtained based
on  the calculation of Wigner (appendix \ref{apendice 3}) and Husimi
functions, using the reduced density matrix (\ref{Bh-Hr density matrix elements}).
We can observe that the entropy start with lower values, this corresponds
to a stage where entanglement and correlation between the states are
weak. According to figure \ref{Bh and Hr density - 1} this happens
in a phase where the quantum states become increasingly squeezed and
displaced, under the influence of the inverted potential. However
when the back reaction begins to grow, correlations and degree of
entanglement between the two states increase, and consequently so
does the entropy, and both subsystems are forced to oscillate (counteracting
the inverted potential). Finally, both states return to their initial
configurations, which brings a reduction of their entropies. It is
in this last phase that, with a decreasing entropy, the information
contained in the state describing Hawking radiation increases as expected
from the Page curve.

At this point, we can ask ourselves: how much the estimate of the
$S_{W}\left(\rho\right)$ give us an accurate description of the real
behaviour of the Von Neumann entropy $S_{VN}\left(\rho\right)$? Since
Wehrl entropy satisfies $S_{W}\left(\rho\right)\geq S_{VN}\left(\rho\right)$,
inspection of figure \ref{Bh and Hr density - 1} tell us that the
variation from a lower values of the entropy (initial stage of the
time evolution) to higher (intermediate stage of the time evolution)
and again to lower values (final stage of the time evolution) seems
to indicate, with reasonable chance, that Von Neumann entropy can
present a behaviour relatively close to Wehrl entropy. In addition,
the fact that we have considered the unitary evolution of the pure
state $\left|\psi_{x_{0}}^{\alpha}\right\rangle \otimes\left|\psi_{y_{0}}^{H}\right\rangle $,
implies that the system matrix density remains a pure state $\left(S_{VN}\left(\rho_{AB}\right)=0\right)$,
while the reduced density matrices, for the two subsystems, correspond
to mixed states $\left(S_{VN}\left(\rho_{A}\right)\neq0\right)$.

\section{Conclusions}
\label{section7}
Even though the simplified model, discussed in this paper, was based
on modest assumptions (namely about the initial black hole quantum
state, among others), it provides a simple mechanism where one can
appreciate the temporal evolution of entropy and the behaviour of
information (in a classical approach with Wehrl entropy being evaluated).
The model has the advantage that it can be treated analytically and
show how the coupling of a harmonic and inverted oscillators can produce
results suggesting how the Page curve can emerge. 

There are certainly many ways in which this model can become more
realistic. However, it would also certainly no longer be able to be
treated analytically, which would inevitably deprive it of its pedagogical
appeal. In one hand, questions such as,
\begin{itemize}
\item how the squeezing parameter evolves in this model?
\item what is the exact behaviour, in this model, of Von Neumann entropy
$S_{VN}\left(\rho\right)$?
\item which aspects of the discussed estimates would benefit by considering
a more realistic model? 
\item how to apply the same procedure to the functional Schrödinger type
of equation (\ref{semi Schr equation})? 
\end{itemize}
can be pursued as possible future topics of investigation. On the
other hand, one can also try to understand to which extent entropy,
and Hawking radiation information, estimates can be made in gravitational
back reaction scenarios such as those proposed in 
\cite{tHooft:1996rdg,tHooft:2018waj}. In that proposal, it is assumed
that particles moving at high speeds to and from the event horizon
cause a drag \cite{DRAY1985173} which has gravitational effects that
can be described by the Aechelburg-Sexl metric \cite{Bonnor1969,Aichelburg1971}.
It is worth to mention that the discussion of back reaction effects of the Hawking radiation and the correct way to derive the Page curve has been an active field of research in connection with the black hole information paradox. The reader can find complete reviews of the problem and recent progresses in that direction in \citep{wang2020nonequilibrium,Marolf_2021,Gautason_2020,Almheiri_2020} 
Finally, the black hole evaporation subject and the fate of the information
enclosed inside it, are crucial aspects that any quantum gravity theory
candidate will have to unveil. At a time when the first direct evidences
of objects that in everything resemble what in General Relativity
is described as a black hole are emerging, our scepticism about their
real existence starts to fade away. However, it has been a long time
since the conceptual problems associated with these hypothetical strange
objects have challenged the limits of theoretical physics.



\vspace{6pt} 

\acknowledgments 
This research work was funded by Funda\c{c}\~ao para a Ci\^encia e a Tecnologia grant number UIDB/MAT/00212/2020. 
\appendix
\section{Wave function time evolution}

\label{apendice 2} In this appendix, we explicitly write the analytical
expressions, for the computation of equation (\ref{Psi Q evolution}),
and the various time functions which help to define state (\ref{Psi X evolution}).
Although some of the following expressions were originally presented in \citep{Kiefer:2008av}, a re-organisation, and introduction of new time functions, used to write equation (\ref{Psi X evolution}) justify the necessity to provide the reader with their accurate modifications.

\subsection{Wave function in the new coordinates}

When the initial state $\left|\psi_{x_{0}}^{\alpha}\right\rangle \otimes\left|\psi_{y_{0}}^{H}\right\rangle $
evolves in time, it defines a wave function that in variables $\left(Q_{1},\,Q_{2}\right)$
is,

\begin{align}
\Psi(Q_{1},\:Q_{2},\:t)= & \left(\frac{m_{\mathrm{Pl}}\omega_{x}}{\pi\hbar}\right)^{1/4}\left(\frac{m_{y}\omega_{y}}{\hbar}\coth\left(\frac{2\pi\omega_{y}\mathrm{GM}}{c^{3}}+i\omega_{y}t_{0}\right)\right)^{1/4}\left(-\frac{\Omega_{1}\Omega_{2}}{\mathcal{F}_{1}\mathcal{F}_{3}}\right)^{1/2}\nonumber \\
 & \exp\left[-\left(\frac{Q_{1}^{2}}{2\hbar}\frac{\mathcal{F}_{2}}{\mathcal{F}_{1}}+\frac{Q_{2}^{2}}{2\hbar}\frac{\mathcal{F}_{4}}{\mathcal{F}_{3}}\right)\right]\nonumber \\
 & \exp\left[-i\alpha^{*}\sqrt{\frac{2\widetilde{\omega}_{x}}{\hbar}}\left(\frac{\Omega_{1}Q_{1}\cos\theta}{\mathcal{F}_{1}}+\frac{\Omega_{2}Q_{2}\sin\theta}{\mathcal{F}_{3}}\right)\right]\nonumber \\
 & \exp\left[\frac{\Omega_{2}Q_{2}\sin2\theta}{2\mathcal{F}_{1}\mathcal{F}_{3}}\left(\frac{\Omega_{1}Q_{1}}{\hbar}-i\alpha^{*}\sqrt{\frac{2\widetilde{\omega}_{x}}{\hbar}}\sinh\Omega_{1}t\:\cos\theta\right)\left(\widetilde{\omega}_{y}+\widetilde{\omega}_{x}\right)\right]\nonumber \\
 & \exp\left[-\frac{\hbar\sin^{2}2\theta\:\sin\Omega_{2}t}{8\mathcal{F}_{1}^{2}\mathcal{F}_{3}}\left(i\frac{\Omega_{1}Q_{1}}{\hbar}+\alpha^{*}\sqrt{\frac{2\widetilde{\omega}_{x}}{\hbar}}\sinh\Omega_{1}t\:\cos\theta\right)^{2}\left(\widetilde{\omega}_{y}+\widetilde{\omega}_{x}\right)^{2}\right]\nonumber \\
 & \exp\left[\alpha^{*2}\widetilde{\omega}_{x}\left(\frac{\cos^{2}\theta\:\sinh\Omega_{1}t}{\mathcal{F}_{1}}+\frac{\sin^{2}\theta\:\sin\Omega_{2}t}{\mathcal{F}_{3}}\right)-\frac{\alpha^{*2}}{2}-\frac{|\alpha|^{2}}{2}\right],\label{WaveFunv Q}
\end{align}
where
\begin{equation}
\widetilde{\omega}_{x}=\frac{\omega_{x}}{\cos^{2}2\theta}\qquad\widetilde{\omega}_{y}=\frac{\omega_{y}}{\cos^{2}2\theta}\coth\left(\frac{2\pi\omega_{y}\mathrm{GM}}{c^{3}}+i\omega_{y}t_{0}\right)
\end{equation}
\begin{align}
\mathcal{F}_{1}= & -i\Omega_{1}\cosh\Omega_{1}t+\widetilde{\omega}_{x}\cos^{2}\theta\sinh\Omega_{1}t+\widetilde{\omega}_{y}\sin^{2}\theta\sinh\Omega_{1}t,\nonumber \\
\mathcal{F}_{2}= & -\Omega_{1}^{2}\sinh\Omega_{1}t-i\Omega_{1}\widetilde{\omega}_{x}\cos^{2}\theta\cosh\Omega_{1}t-i\Omega_{1}\widetilde{\omega}_{y}\sin^{2}\theta\cosh\Omega_{1}t,\nonumber \\
\mathcal{F}_{3}= & -i\Omega_{2}\cos\Omega_{2}t+\widetilde{\omega}_{x}\sin^{2}\theta\sin\Omega_{2}t+\widetilde{\omega}_{y}\cos^{2}\theta\sin\Omega_{2}t,\nonumber \\
 & -(\widetilde{\omega}_{y}+\widetilde{\omega}_{x})^{2}\sin^{2}2\theta\frac{\sinh\Omega_{1}t\:\sin\Omega_{2}t}{4\mathcal{F}_{1}}\nonumber \\
\mathcal{F}_{4}= & \Omega_{2}^{2}\sin\Omega_{2}t-i\Omega_{2}\widetilde{\omega}_{x}\sin^{2}\theta\cos\Omega_{2}t-i\Omega_{2}\widetilde{\omega}_{y}\cos^{2}\theta\cos\Omega_{2}t\nonumber \\
 & +i\Omega_{2}(\widetilde{\omega}_{y}+\widetilde{\omega}_{x})^{2}\sin^{2}2\theta\frac{\sinh\Omega_{1}t\:\cos\Omega_{2}t}{4\mathcal{F}_{1}}.\label{TimeFunc1}
\end{align}
When we reverse the coordinate transformation $\Psi(Q_{1},\:Q_{2},\:t)\rightarrow\Psi(x,\:y,\:t)$,
applying transformations (\ref{Q to X coord transform}), we obtain
the state defined in equation (\ref{Psi X evolution}), where, 
\begin{align}
F(t)= & \left(\frac{m_{\mathrm{Pl}}\omega_{x}}{\pi\hbar}\right)^{1/4}\left(\frac{m_{y}\omega_{y}}{\hbar}\coth\left(\frac{2\pi\omega_{y}\mathrm{GM}}{c^{3}}+i\omega_{y}t_{0}\right)\right)^{1/4}\left(-\frac{\Omega_{1}\Omega_{2}}{\mathcal{F}_{1}\mathcal{F}_{3}}\right)^{1/2}\times\nonumber \\
 & \exp\left[\alpha^{*2}\widetilde{\omega}_{x}\left(\frac{\cos^{2}\theta\:\sinh\Omega_{1}t}{\mathcal{F}_{1}}+\frac{\sin^{2}\theta\:\sin\Omega_{2}t}{\mathcal{F}_{3}}\right)-\frac{\alpha^{*2}}{2}-\frac{|\alpha|^{2}}{2}\right]\times\nonumber \\
 & \exp\left[\frac{\alpha^{*2}\widetilde{\omega}_{x}(\widetilde{\omega}_{y}+\widetilde{\omega}_{x})^{2}\cos^{2}\theta\:\sin^{2}2\theta\:\sinh^{2}\Omega_{1}t\:\sin\Omega_{2}t}{4\mathcal{F}_{1}^{2}\mathcal{F}_{3}}\right],\nonumber \\
A(t)= & \dfrac{m_{\mathrm{Pl}}}{2\hbar}\biggl[\frac{\mathcal{F}_{2}}{\mathcal{F}_{1}}\cos^{2}\theta+\frac{\mathcal{F}_{4}}{\mathcal{F}_{3}}\sin^{2}\theta+\frac{\Omega_{1}\Omega_{2}}{2\mathcal{F}_{1}\mathcal{F}_{3}}(\widetilde{\omega}_{y}+\widetilde{\omega}_{x})\sin^{2}2\theta\nonumber \\
 & +\frac{\Omega_{1}^{2}(\widetilde{\omega}_{y}+\widetilde{\omega}_{x})^{2}\sin^{2}2\theta\:\cos^{2}\theta\:\sin\Omega_{2}t}{4\mathcal{F}_{1}^{2}\mathcal{F}_{3}}\biggr],\nonumber \\
B(t)= & -i\alpha^{*}\sqrt{\frac{2m_{\mathrm{Pl}}\widetilde{\omega}_{x}}{\hbar}}\biggl[\frac{\Omega_{1}\cos^{2}\theta}{\mathcal{F}_{1}}-\frac{\Omega_{2}\sin2\theta}{2\mathcal{F}_{3}}-\frac{\Omega_{2}(\widetilde{\omega}_{y}+\widetilde{\omega}_{x})\sin^{2}2\theta\:\sinh\Omega_{1}t}{4\mathcal{F}_{1}\mathcal{F}_{3}}\nonumber \\
 & -\frac{\Omega_{1}(\widetilde{\omega}_{y}+\widetilde{\omega}_{x})^{2}\sin^{2}2\theta\:\cos^{2}\theta\:\sinh\Omega_{1}t\:\sin\Omega_{2}t}{4\mathcal{F}_{1}^{2}\mathcal{F}_{3}}\biggr],\nonumber \\
C(t)= & \dfrac{m_{y}}{2\hbar}\biggl[\frac{\mathcal{F}_{2}}{\mathcal{F}_{1}}\sin^{2}\theta+\frac{\mathcal{F}_{4}}{\mathcal{F}_{3}}\cos^{2}\theta+\frac{\Omega_{1}\Omega_{2}}{2\mathcal{F}_{1}\mathcal{F}_{3}}(\widetilde{\omega}_{y}+\widetilde{\omega}_{x})\sin^{2}2\theta\nonumber \\
 & +\frac{\Omega_{1}^{2}(\widetilde{\omega}_{y}+\widetilde{\omega}_{x})^{2}\sin^{2}2\theta\:\sin^{2}\theta\:\sin\Omega_{2}t}{4\mathcal{F}_{1}^{2}\mathcal{F}_{3}}\biggr],\nonumber \\
D\left(t\right)= & \frac{i\alpha^{*}}{2}\sqrt{\frac{2m_{y}\widetilde{\omega}_{x}}{\hbar}}\sin2\theta\biggl(-\frac{\Omega_{2}}{\mathcal{F}_{1}\mathcal{F}_{3}}(\widetilde{\omega}_{y}+\widetilde{\omega}_{x})\cos^{2}\theta\:\sinh\Omega_{1}t-\frac{\Omega_{2}}{\mathcal{F}_{3}}+\frac{\Omega_{1}}{\mathcal{F}_{1}}\nonumber \\
 & -\frac{\Omega_{1}(\widetilde{\omega}_{y}+\widetilde{\omega}_{x})^{2}\sin^{2}2\theta\:\sinh\Omega_{1}t\:\sin\Omega_{2}t}{4\mathcal{F}_{1}^{2}\mathcal{F}_{3}}\biggr),\nonumber \\
E(t)= & \frac{\sin2\theta}{\hbar}\sqrt{m_{\mathrm{Pl}}m_{y}}\biggl(\frac{\Omega_{1}\Omega_{2}}{2\mathcal{F}_{1}\mathcal{F}_{3}}(\widetilde{\omega}_{y}+\widetilde{\omega}_{x})\cos2\theta+\frac{\mathcal{F}_{2}}{\mathcal{F}_{1}}+\frac{\mathcal{F}_{4}}{\mathcal{F}_{3}}\nonumber \\
 & +\frac{\Omega_{1}^{2}(\widetilde{\omega}_{y}+\widetilde{\omega}_{x})^{2}\sin^{2}2\theta\:\sin\Omega_{2}t}{8\mathcal{F}_{1}^{2}\mathcal{F}_{3}}\biggr)\:.\label{TimeFunc2}
\end{align}


\section{Wigner functions}

\label{apendice 3} In this appendix, we obtain the Wigner functions
analytic expressions for the reduced density matrices (\ref{Bh-Hr density matrix elements}).
In addition, we present the numerical simulations for the Wigner function
time evolution related to the black hole subsystem (moreover, for
the Hawking radiation subsystem the simulations display some similarity).
The main idea is also to illustrate the effects of the displacement
$\hat{D}\left(\alpha\right)$ and squeeze $\hat{S}\left(\xi\right)$
operators, while Wigner function evolves in time in phase space. 

\subsection{Black hole and Hawking radiation Wigner functions}

\begin{figure}[ht]
\begin{centering}
\includegraphics[width=6cm]{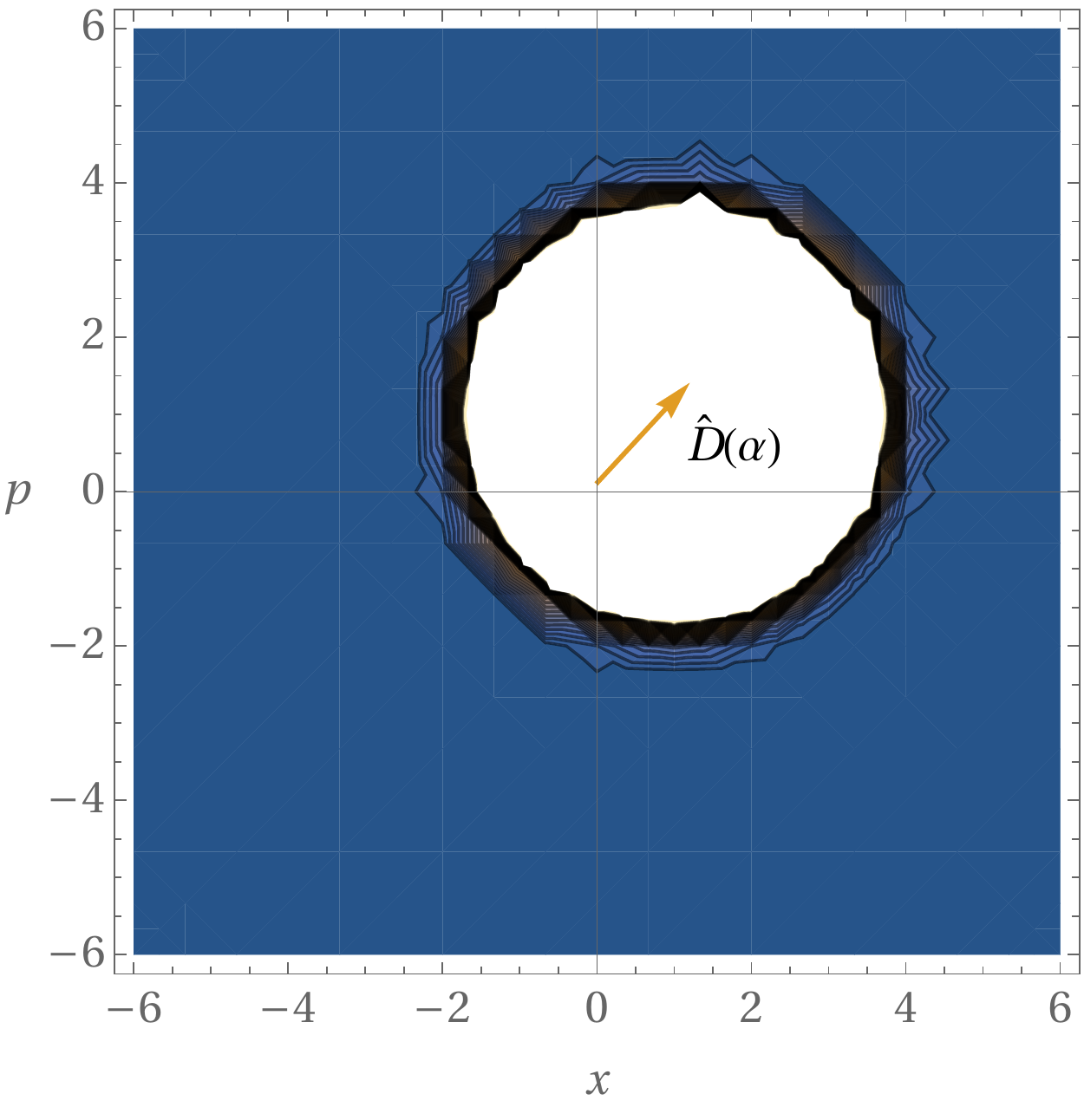}\quad{}\includegraphics[width=6cm]{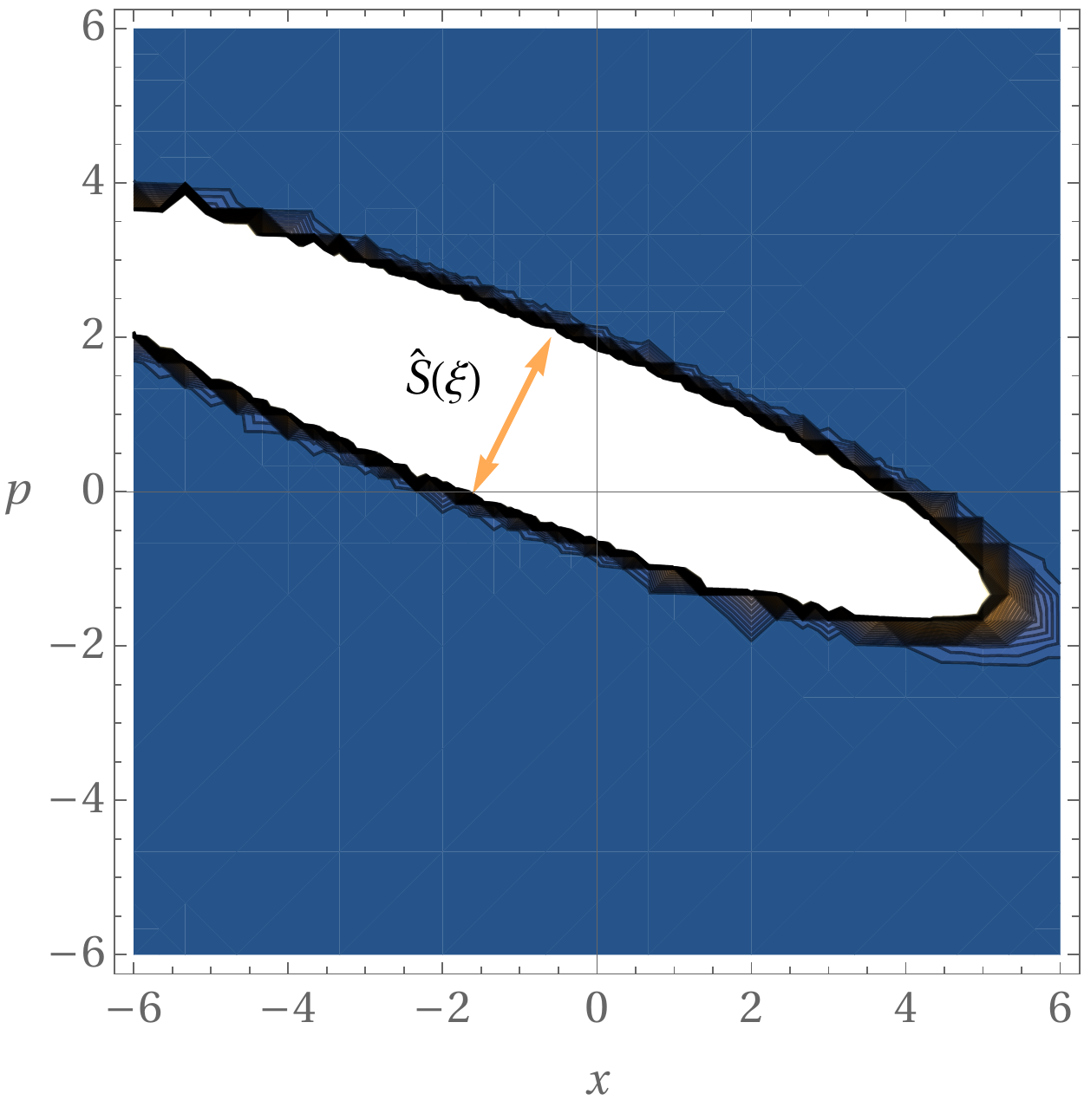}
\par\end{centering}
\begin{centering}
\includegraphics[width=6cm]{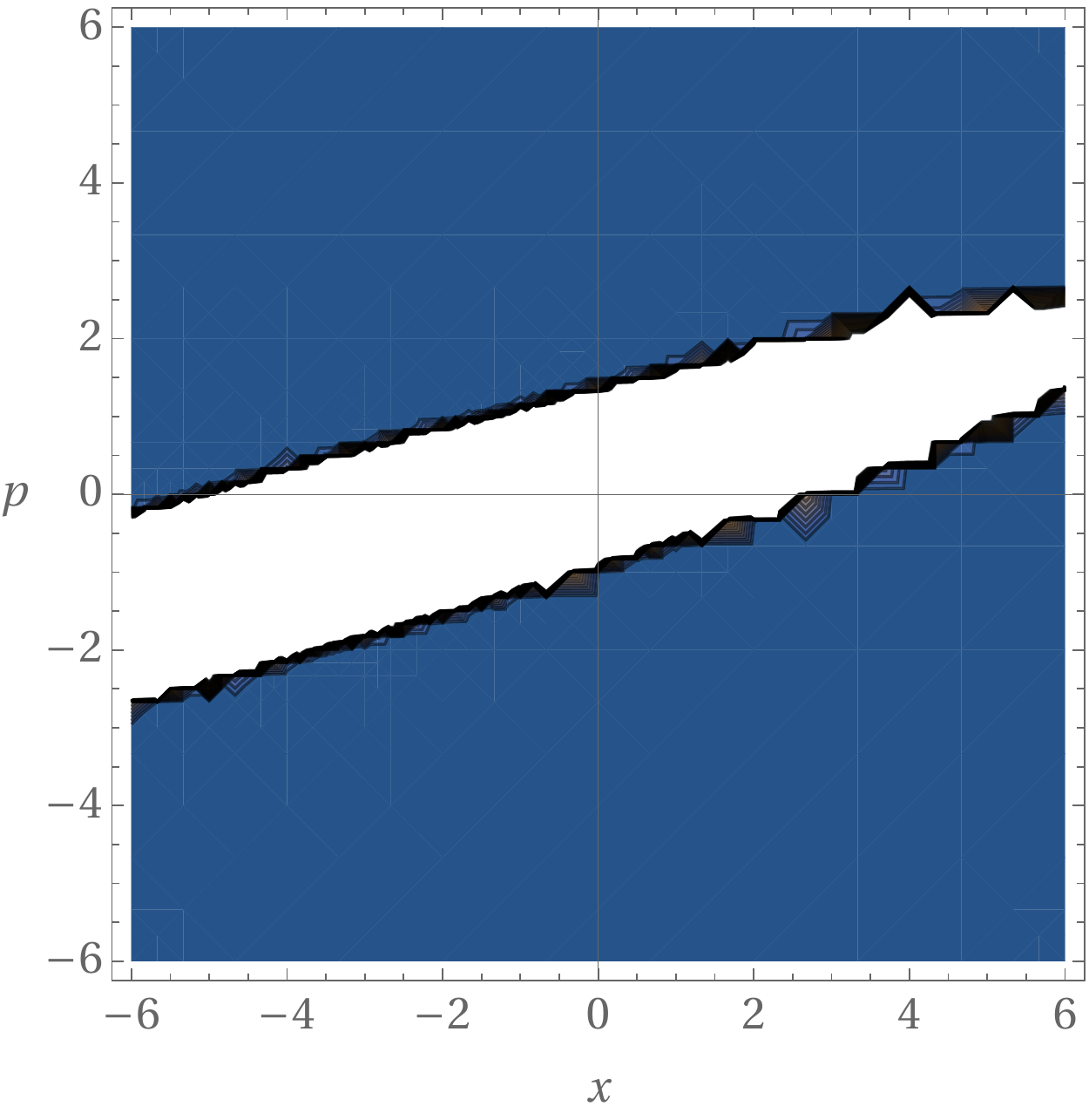}\quad{}\includegraphics[width=6cm]{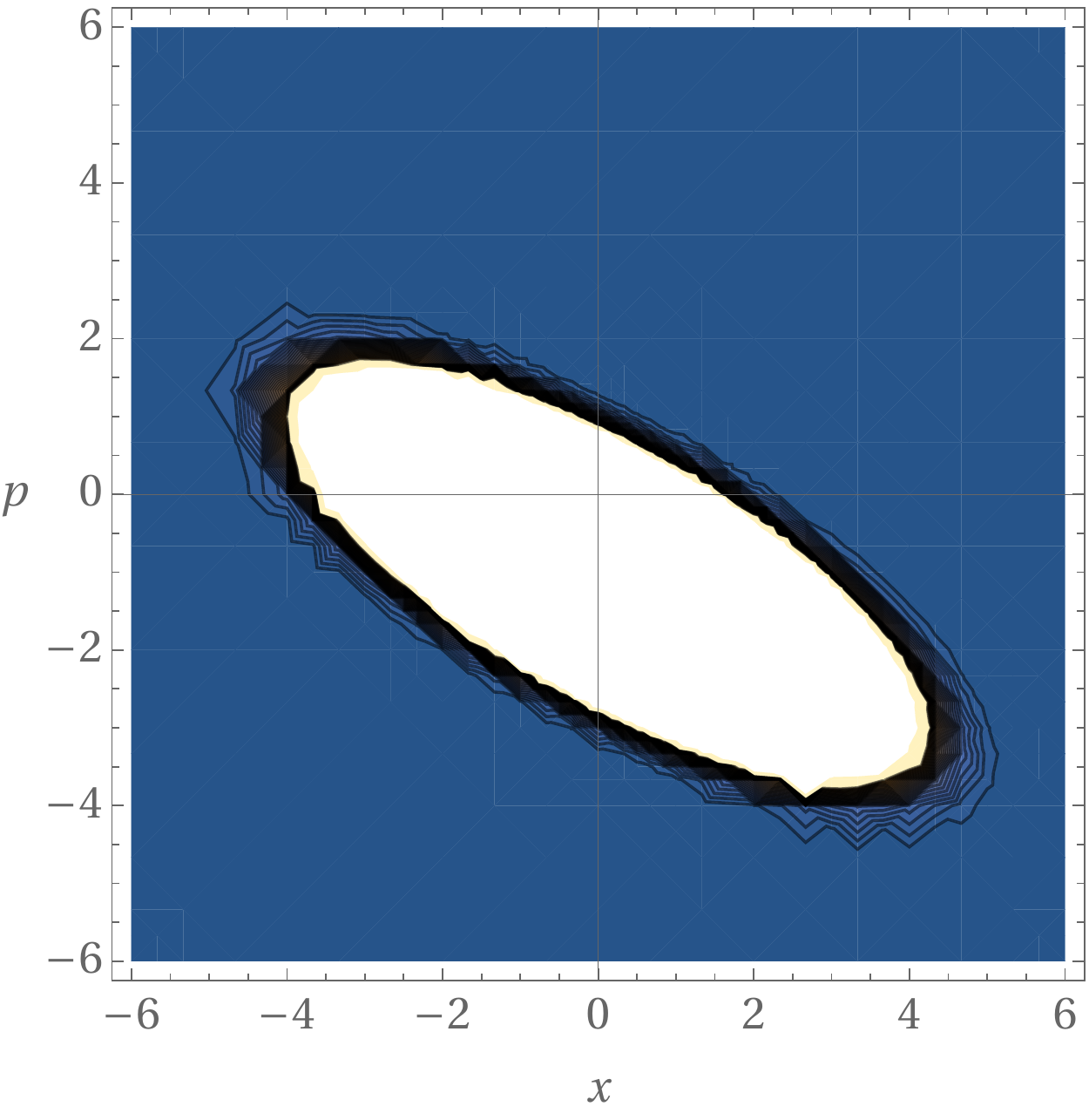}
\par\end{centering}
\begin{centering}
\includegraphics[width=6cm]{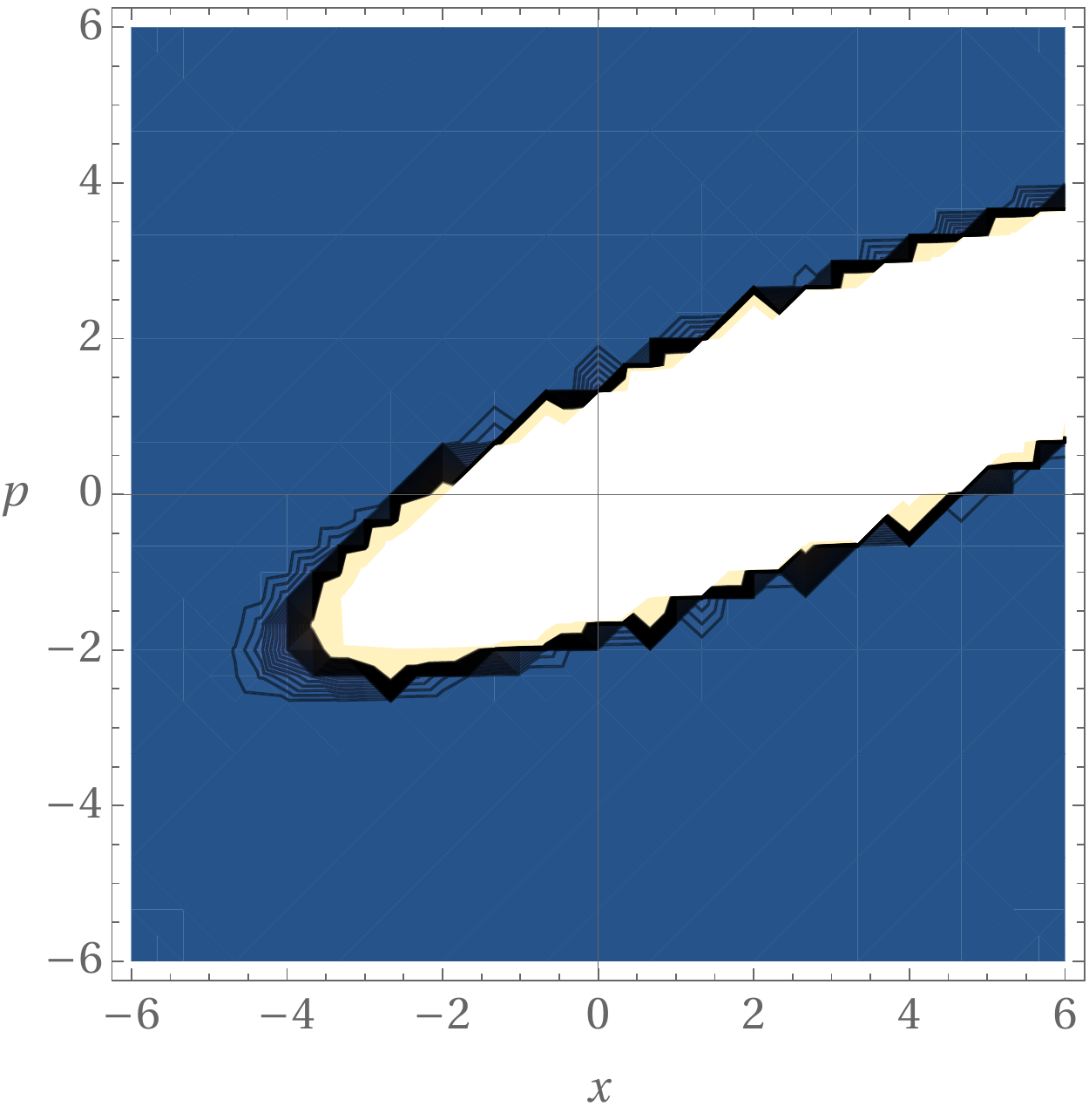}\quad{}\includegraphics[width=6cm]{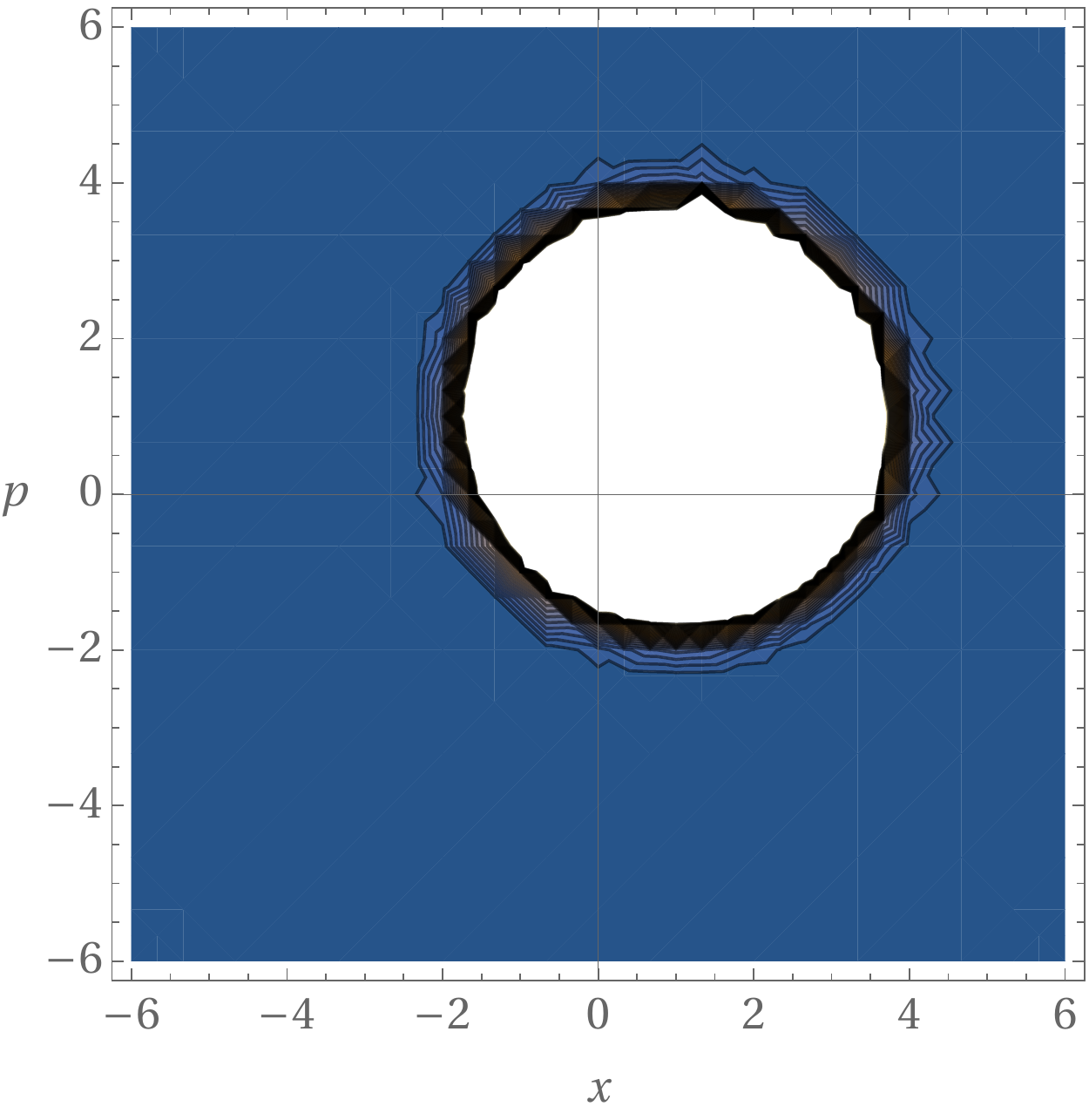}
\par\end{centering}
\centering{}\caption{{\footnotesize{}\label{ Wigner plot1} Time evolution of the black
hole state Wigner function, with $m_{\mathrm{Pl}}=\hbar=\omega_{x}=x_{0}=1$;
$p_{0}=-1$. The coupling parameter which defines the
back reaction is $\mu=1.01$, with $\omega_{y}=\omega_{x}\times10^{5/2}$
and $m_{y}=m_{\mathrm{Pl}}\times10^{-5}$. We verify that, throughout
the various stages of the evolution (corresponding to the various
panels), the action of the operators $\hat{D}\left(\alpha\right)$
(displacement operator) and $\hat{S}\left(\xi\right)$ (squeeze operator),
produces a full rotation of the displacement center point of the initial
Wigner function around the origin of phase space, while various degrees
of squeezing affect the shape of the state.}}
\end{figure}

The Wigner function can be defined as, 

\begin{equation}
W_{Bh}\left(x,\,p\right)=\dfrac{1}{\pi\hbar}\int d\eta\:\dfrac{1}{\pi\hbar}\int d\eta\:e^{-ip\eta/\hbar}\rho_{Bh}\left(x+\dfrac{\eta}{2},\,x-\dfrac{\eta}{2}\right)\:,
\end{equation}
where, upon the substitution of $\rho_{Bh}$ by equation (\ref{Bh-Hr density matrix elements}),
we get
\begin{equation}
\begin{aligned}W_{Bh}\left(x,\,p\right) & =\dfrac{1}{\pi\hbar}\int d\eta\:e^{-ip\eta/\hbar}\rho_{Bh}\left(x+\dfrac{\eta}{2},\,x-\dfrac{\eta}{2}\right)\\
 & =\dfrac{\mathcal{N}_{1}}{\pi\hbar}\int d\eta\:e^{-ip\eta/\hbar}\exp\Biggl(-\mathcal{A}_{1}\left(x-\dfrac{\eta}{2}\right)^{2}+\mathcal{B}_{1}\left(x-\dfrac{\eta}{2}\right)\\
 & \qquad\qquad\qquad\qquad-\mathcal{A}_{1}^{*}\left(x+\dfrac{\eta}{2}\right)^{2}+\mathcal{B}_{1}^{*}\left(x+\dfrac{\eta}{2}\right)+\left|\mathcal{C}_{1}\right|\left(x^{2}-\dfrac{\eta^{2}}{4}\right)\Biggr)
\end{aligned}
\:.
\end{equation}
After some algebraic manipulation, we obtain
\begin{equation}
\begin{aligned}W_{Bh}\left(x,\,p\right)= & \dfrac{\mathcal{N}_{1}}{\pi\hbar}\exp\left(-\left(2\mathrm{Re\left(\mathcal{A}_{1}\right)-}\left|\mathcal{C}_{1}\right|\right)x^{2}+2\mathrm{Re\left(\mathcal{B}_{1}\right)x}\right)\cdot\\
 & \cdot\int d\eta\:\exp\left(-\left(2\mathrm{Re\left(\mathcal{A}_{1}\right)-}\left|\mathcal{C}_{1}\right|\right)\dfrac{\eta^{2}}{4}+i\left(2\mathrm{Im}\left(\mathcal{A}_{1}\right)x-\mathrm{Im}\left(\mathcal{B}_{1}\right)-\dfrac{p}{\hbar}\right)\eta\right)
\end{aligned}
\:,
\end{equation}
\begin{equation}
\begin{aligned}W_{Bh}\left(x,\,p\right)= & \dfrac{2\mathcal{N}_{1}}{\sqrt{\pi}\hbar\sqrt{2\mathrm{Re\left(\mathcal{A}_{1}\right)-}\left|\mathcal{C}_{1}\right|}}\exp\left(-\left(2\mathrm{Re\left(\mathcal{A}_{1}\right)-}\left|\mathcal{C}_{1}\right|\right)x^{2}+2\mathrm{Re\left(\mathcal{B}_{1}\right)}x\right)\cdot\\
 & \qquad\qquad\qquad\qquad\quad\cdot\exp\left(-\frac{(2\mathrm{Im}\left(\mathcal{A}_{1}\right)x-\mathrm{Im}\left(\mathcal{B}_{1}\right)-p/\hbar)^{2}}{(2\mathrm{Re\left(\mathcal{A}_{1}\right)-}\left|\mathcal{C}_{1}\right|)}\right)
\end{aligned}
\:,\label{Bh Wigner}
\end{equation}
where
\begin{equation}
\begin{cases}
\mathcal{A}_{1}= & A(t)-\dfrac{E^{2}\left(t\right)}{8\mathrm{Re}\left(C\left(t\right)\right)}\\
\mathcal{B}_{1}= & B(t)+\dfrac{\mathrm{Re}\left(D\left(t\right)\right)}{\mathrm{Re}\left(C\left(t\right)\right)}\dfrac{E\left(t\right)}{2}\\
\left|\mathcal{C}_{1}\right|= & \dfrac{\left|E\left(t\right)\right|^{2}}{4\mathrm{Re}\left(C\left(t\right)\right)}\\
\mathcal{N}_{1}= & \left|F\left(t\right)\right|^{2}\exp\left(\dfrac{\mathrm{Re}\left(D\left(t\right)\right)}{2\mathrm{Re}\left(C\left(t\right)\right)}\right)\sqrt{\dfrac{\pi\hbar}{m_{y}\mathrm{Re}\left(C\left(t\right)\right)}}
\end{cases}\label{Bh Wigner coef}
\end{equation}
Concerning the Hawking radiation, a similar procedure enable us to
obtain the following Wigner function,
\begin{equation}
\begin{aligned}W_{Hr}\left(y,\,p\right)= & \dfrac{2\mathcal{N}_{2}}{\sqrt{\pi}\hbar\sqrt{2\mathrm{Re\left(\mathcal{A}_{2}\right)-}\left|\mathcal{C}_{2}\right|}}\exp\left(-\left(2\mathrm{Re\left(\mathcal{A}_{2}\right)-}\left|\mathcal{C}_{2}\right|\right)y^{2}+2\mathrm{Re\left(\mathcal{B}_{2}\right)}y\right)\cdot\\
 & \qquad\qquad\qquad\qquad\quad\cdot\exp\left(-\frac{(2\mathrm{Im}\left(\mathcal{A}_{2}\right)y-\mathrm{Im}\left(\mathcal{B}_{2}\right)-p/\hbar)^{2}}{(2\mathrm{Re\left(\mathcal{A}_{2}\right)-}\left|\mathcal{C}_{2}\right|)}\right)
\end{aligned}
\label{Hr Wigner}
\end{equation}
where
\begin{equation}
\begin{cases}
\mathcal{A}_{2}= & C(t)-\dfrac{E^{2}\left(t\right)}{8\mathrm{Re}\left(A\left(t\right)\right)}\\
\mathcal{B}_{2}= & D(t)+\dfrac{\mathrm{Re}\left(B\left(t\right)\right)}{\mathrm{Re}\left(A\left(t\right)\right)}\dfrac{E\left(t\right)}{2}\\
\left|\mathcal{C}_{2}\right|= & \dfrac{\left|E\left(t\right)\right|^{2}}{4\mathrm{Re}\left(A\left(t\right)\right)}\\
\mathcal{N}_{2}= & \left|F\left(t\right)\right|^{2}\exp\left(\dfrac{\mathrm{Re}\left(B\left(t\right)\right)}{2\mathrm{Re}\left(A\left(t\right)\right)}\right)\sqrt{\dfrac{\pi\hbar}{m_{y}\mathrm{Re}\left(A\left(t\right)\right)}}
\end{cases}\label{Hr Wigner coef}
\end{equation}
In figure \ref{ Wigner plot1} we display the time evolution for function
$W_{Bh}\left(x,\,p\right)$ in the interval $t\sim\left[0,\,40\right]$,
which is related to figure \ref{Bh and Hr density - 1}. This time
interval can approximately be taken as measuring one full cycle of
`oscillation' for the black hole state, i. e. , the average time
required for the state to return to its initial configuration. Inspecting
the aforementioned figure, we can notice that the initial state Wigner
function (first left panel of the figure) describes a coherent state
$\left|\psi_{x_{0}}^{\alpha}\right\rangle $, which is displaced from
the origin of phase space, since 
\begin{equation}
\left|\psi_{x_{0}}^{\alpha}\right\rangle =\hat{D}\left(\alpha\right)\left|0\right\rangle \:,
\end{equation}
in agreement with equation (\ref{Desloc estado coerente}). After
some time has elapsed (top right panel of the figure), the Wigner
function starts to squeeze, in the density plot, deforming its initial
circular shape to an elliptical one. This illustrates the action of
the squeeze operator $\hat{S}\left(\xi\right)$, besides the displacement
around the origin of phase space. Finally, we can observe that a full
rotation of the displacement center point occurs around the origin
of phase space, while various degrees of squeezing affect the shape
of the state.

\bibliographystyle{utphys}
\bibliography{References}

\end{document}